\documentclass[%
  aip,
   jmp,%
  amsmath,amssymb,
 preprint,%
 ]{revtex4-1}

\usepackage{graphicx}
\usepackage{dcolumn}
\usepackage{bm}

\usepackage{graphicx}
\usepackage{dcolumn}
\usepackage{bm}
\usepackage{amssymb}  
\usepackage{commath}
\usepackage{amsmath}

\usepackage{amssymb}
\usepackage{amsmath}
\usepackage{commath}
\usepackage{graphicx,bm}
\usepackage{verbatim}
\usepackage{rotating}
\PassOptionsToPackage{export}{adjustbox}
\usepackage[export]{adjustbox}
\usepackage{adjustbox}

\usepackage{graphicx}
\usepackage[utf8]{inputenc}
\usepackage{wrapfig}
\usepackage{siunitx}

\usepackage{placeins}
\usepackage{float}

\usepackage{soul}
\usepackage{xcolor}

\usepackage[utf8]{inputenc}

\usepackage{amssymb}
\usepackage{pifont}
\newcommand{\cmark}{\ding{51}}%
\newcommand{\xmark}{\ding{55}}%

 \usepackage{multirow}
 \definecolor{orange}{rgb}{1,0.5,0}
\definecolor{vc}{rgb}{0.85, 0.11, 0.51}

\begin{document}


\title[]{An analytic virtual-source-based current-voltage model for ultra-thin black phosphorus field-effect transistors}

\author{Elahe Yarmoghaddam}
\email{ey651@nyu.edu.}
\affiliation{ 
Electrical and Computer Engineering Department, New York University, New York, NY 11201, USA}


\author{Nazila Haratipour}

\affiliation{ 
Department of Electrical and Computer Engineering, University of Minnesota, Minneapolis, MN 55455, USA}

\author{Steven J. Koester}
\email{skoester@umn.edu.}

\affiliation{ 
Department of Electrical and Computer Engineering, University of Minnesota, Minneapolis, MN 55455, USA}

\author{Shaloo Rakheja}%
\email{shaloo.rakheja@nyu.edu.}
\affiliation{ 
Electrical and Computer Engineering Department, New York University, New York, NY 11201, USA}

\begin{abstract}
In this paper, we develop an analytic physics-based 
model to describe current conduction in 
ultra-thin black phosphorus (BP) field-effect transistors (FETs). The model extends the concept of virtual source charge calculation to capture the effect of both hole and electron charges for ambipolar transport characteristics.
The model comprehends the in-plane band-structure anisotropy in BP, as well as the asymmetry in electron and hole current conduction characteristics.
The model also includes the effect of Schottky-type source/drain contact resistances, which are voltage-dependent and can significantly limit current conduction in the on-state in BP FETs.
Model parameters are extracted using measured data of 
back-gated BP transistors with gate lengths of 1000 nm and 300 nm with BP thickness of 7.3 nm and 8.1 nm, and for the temperature range of 180 K to 298 K.
Compared to previous BP models that are validated only for room-temperature and near-equilibrium bias conditions (low drain-source voltage), we demonstrate excellent agreement between the data and model over a broad range of bias and temperature values.
The model is also validated against numerical TCAD data of top-gated BP transistors with a channel length of 300 nm. The model is implemented in Verilog-A and the capability of the model to handle both dc and transient circuit simulations is demonstrated using SPECTRE.
The model not only provides a physical insight into technology-device interaction in BP transistors, but can also be used to design and optimize BP-based circuits using a standard hierarchical circuit simulator.

\end{abstract}


\maketitle

\vspace{-10pt}
\section{\label{sec:level1}Introduction}
\vspace{-10pt}
Over the past decade, two dimensional (2D) materials, including graphene,\cite{yarmoghaddam2017dispersion, rakheja2013evaluation, novoselov2005two} transition metal dichalcogenides (TMDC),\cite{wang2012electronics, fivaz1967mobility} silicene,\cite{houssa2011electronic, vogt2012silicene} and germanane, \cite{bianco2013stability} have emerged as promising candidates for future generations of nanoelectronic devices due to their excellent electrostatic integrity, vertical scalability, 
and electronic properties that are
considerably different from those in their bulk parental materials.~\cite{xia2014two,das2014ambipolar,yi2017two, qiao2014high} 
In graphene, carriers have a limited phase space for scattering, which can allow quasi-ballistic transport at room-temperature.~\cite{geim2007rise, hwang2008acoustic, qiao2014high} Several high-frequency devices using graphene have been experimentally demonstrated such as 
radio-frequency FETs, optical modulators, and photo-detectors.~\cite{xia2014two} 
However, the use of graphene in digital switching applications is challenging due to its low on-off current ratio, which results from its zero band gap.~\cite{han2014strain} 
On the other hand, TMDC-based FETs, with the
band gap in the range of ($1-2$) eV, possess high on-off current ratios;\cite{qiao2014high} however, the carrier mobility in TMDCs
is much lower than that in graphene.\cite{du2014device}

More recently,
black phosphorus (BP) has emerged as one of the most interesting 2D materials for high performance transistor applications.\cite{huang2017black, haratipour2016fundamental} 
In its bulk form, BP exhibits a
band gap $\approx$ 0.3 eV, while in its monolayer form, the band
gap increases to $\approx$ 2 eV.~\cite{haratipour2016fundamental, xia2014two, yang2016few} 
Additionally, the hole mobility in BP is reported to be as high as 1000 {cm$^2$/Vs} at room-temperature.~\cite{bohloul2016theoretical, haratipour2016fundamental} Because of its puckered crystal structure, there exists a significant difference in the effective mass of carriers along the zigzag and armchair directions, with the armchair direction featuring light mass of carriers.~\cite{haratipour2016fundamental, luo2015anisotropic, xia2014two, yang2016few} 
With its tunable band
gap, band-structure anisotropy, and high carrier mobility, BP 
is an excellent material to implement digital transistors for both high-performance and low-power electronic applications.~\cite{xia2014rediscovering, yang2016few}

To understand, design, and simulate electronic circuits built with BP transistors, a physics-based compact model of current
conduction is needed.\cite{rakheja2014ambipolar} 
So far, only a few models that describe current conduction in BP transistors have been reported in the literature \cite{penumatcha2015analysing, esqueda2017transport}. Prior BP FET models are mainly suited for near-equilibrium transport conditions, i.e. when the drain-source bias, $V_\mathrm{ds}$, is comparable to the thermal voltage, $\phi_t = k_BT/q$ ($k_B$ is Boltzmann constant, $T$ is the operating temperature, and $q$ is the elementary charge).   
Moreover, these models also do not include the effect of interface traps and non-linear contact resistances, which are important to interpret experimental data. 
The 2D FET models reported elsewhere either introduce significant empiricism in their approach due to thermally activated and hopping-based transport,~\cite{wang2016surface, wang2017unified, cao2018new} or suffer from a limitation of not being able to reproduce the ambipolar characteristics.
For example, the current-voltage (I-V) model for short-channel 2D transistors presented in Ref.~\citenum{taur2016short} focuses mainly on intrinsic current conduction, while extrinsic effects due to contacts and traps are not included. 
The drift-diffusive I-V model presented in Ref.~\citenum{marin2018new} is based on the calculation of the surface potential throughout the channel. However, the desired error-tolerance in the calculation of the surface potential must be in the range of sub-nV, which makes this model susceptible to convergence issues. Additionally, the broad-bias validity of the model in Ref.~\onlinecite{marin2018new} requires numerical integration, which increases the computational complexity of the proposed model. As such, 
the model fit to experimental data has been demonstrated only for a limited bias range (gate-source bias greater than -1.5 V for transfer characteristics and drain-source bias of -1.2 V for output characteristics). Models based on the Landauer transport theory~\cite{landauer1957spatial, datta1997electronic} have also been reported in the literature.~\cite{penumatcha2015analysing, esqueda2017transport} In Ref.~\citenum{penumatcha2015analysing}, a Schottky-barrier (SB) model to describe current conduction in off-state is developed.
This model is extended in Ref.~\citenum{esqueda2017transport} to cover the on-state regime by including the channel transmission. 
However, this model is not suited for broad-bias circuit simulations as its validity is restricted to $V_\mathrm{ds} \leq$ 10 mV at 300 K. We also note that none of the existing compact models for BP transistors has been implemented in Verilog-A to enable device-circuit co-design and optimization.

In this paper, we present an analytic I-V model of BP transistors to capture the ambipolar nature of current conduction over a broad range of bias voltages and temperatures. 
This model is based on the calculation of channel charges at the virtual source and is an extension of the previously published ambipolar virtual-source model applicable for graphene FETs.~\cite{rakheja2014ambipolar}
This paper extends the model in Ref.~\citenum{rakheja2014ambipolar} in the following key ways. 
First, the threshold voltage, $V_\mathrm{t}$, for electron and hole conduction is redefined due to the FET structure under study. Second, to handle the nonlinear behavior of Schottky source/drain contacts, we develop a new contact resistance model. Third, the model is extended to capture the low-temperature behavior of BP FETs. The model is validated by applying it to study BP FETs with gate lengths of 1000 nm and 300 nm and with BP thickness of 7.3 nm and 8.1 nm. 
This model has been implemented in Verilog-A and is used to simulate the circuit behavior of BP-based inverters and ring oscillators.

\vspace{-10pt}
\section{Model Description}
\vspace{-10pt}
In transistors with ambipolar current conduction, the net drain-source current, $I_\mathrm{ds}$, for a given device width, $W$, is given as~\cite{rakheja2014ambipolar}
\begin{subequations}
\begin{equation}
\vspace{-10pt}
    I_\mathrm{ds} = I_\mathrm{elec} + I_\mathrm{hole},
\end{equation}
\vspace{-10pt}
\begin{equation}
    I_\mathrm{elec} = WQ_{\mathrm{x0},e}v_{\mathrm{x0},e}F_{\mathrm{sat},e},
\end{equation}
\vspace{-10pt}
\begin{equation}
    I_\mathrm{hole} = WQ_{\mathrm{x0},h}v_{\mathrm{x0},h}F_{\mathrm{sat},h}.
\end{equation}
\label{eq:net_I}
\end{subequations}
Note all quantities calculated at the top-of-the-barrier or the virtual source (VS) are denoted with the subscript x$_0$. 
In the above equation, $Q_{\mathrm{x0},e}$ and $Q_{\mathrm{x0},h}$ are the electron and hole charges, respectively, at the VS point in the channel (see Section~\ref{sec:Qmodel} for details.) The velocities, $v_{\mathrm{x0},e}$ and $v_{\mathrm{x0,}h}$, are the electron and hole saturation velocities, respectively. The functions, $F_{\mathrm{sat},e}$ and $F_{\mathrm{sat},h}$ empirically capture the transition between linear and saturation regimes for the electron and hole branches, respectively. Per this model, it is assumed that there exist two VS points at opposite ends in the channel for electrons and holes. Unlike graphene FETs in which electron and holes typically have similar mobilities and velocities, BP FETs feature different physical properties for electrons and holes. As a result of this difference, $v_{\mathrm{x0},e}$ ($F_{\mathrm{sat},e}$) and $v_{\mathrm{x0},h}$ ($F_{\mathrm{sat},h}$) are treated separately in this paper. Moreover, in current state-of-the-art technology, the carrier mean free paths in BP FETs are on the order of few 10's of nanometers. Given that the devices under study have gate lengths on the order of several 100's of nanometers, we expect transport to be collision dominated \cite{lundstrom2014compact}. As such, the velocities in Eq.~(\ref{eq:net_I}) are treated as saturation velocities, rather than as injection velocities in the quasi-ballistic transport. This modified interpretation of $v_\mathrm{x0}$ is similar to that reported in Ref.~\citenum{radhakrishna2014virtual} to handle transport in long-channel gallium nitride transistors using the VS model.

The empirical function $F_{\mathrm{sat},i}$ ($i$ = $e/h$ for electrons/holes) is given as \cite{rakheja2014ambipolar}
\begin{equation}
    F_{\mathrm{sat},i} = \frac{V_\mathrm{dsi}/V_{\mathrm{dsat},i}}{\left(1+\left(V_\mathrm{dsi}/V_{\mathrm{dsat},i}\right)^{\beta_i}\right)^{1/\beta_i}}.
\end{equation}
In this function, $V_\mathrm{dsi}$ is the intrinsic drain-source voltage drop in the channel, $\beta_i$ is an empirical parameter, typically in the range of 1.5 to 2.5, and is obtained upon
calibration with experimental data, and $V_{\mathrm{dsat},i} = v_{\mathrm{x0},i}L/\mu_i$ is the drain-source voltage at which the current conduction changes from linear to saturation regimes. Here, $L$ is the channel length and $\mu_i$ is the mobility of carriers. Details of mobility and its temperature dependence are presented in Section~\ref{sec:temp}.

All quantities in Eq.~(\ref{eq:net_I}) vary with $V_\mathrm{gsi}$ and $V_\mathrm{dsi}$, which are the intrinsic gate-source and drain-source voltages, respectively. Intrinsic voltages are given as~\cite{rakheja2014ambipolar}
\begin{subequations}
\begin{equation}
V_\mathrm{gsi} = V_\mathrm{gs} - \left(I_\mathrm{elec}R_\mathrm{elec} + I_\mathrm{hole}R_\mathrm{hole}\right)
\label{eq:Vi},
\end{equation}
\begin{equation}
V_\mathrm{dsi} = V_\mathrm{ds} - 2\left(I_\mathrm{elec}R_\mathrm{elec} + I_\mathrm{hole}R_\mathrm{hole}\right),
\end{equation}
\end{subequations}
\hspace{-1pt}where $V_\mathrm{gs}$ and $V_\mathrm{ds}$ denote the external gate-source and drain-source voltage drops, respectively. Contact resistances corresponding to electron and hole branches are denoted as $R_\mathrm{elec}$ and $R_\mathrm{hole}$, respectively. Section~\ref{sec:contact} presents the analytic model of contact resistances.

\vspace{-10pt}
\subsection{Channel charge model} \label{sec:Qmodel}
\vspace{-10pt}
The analytic model of electron and hole charges in Eq.~(\ref{eq:net_I}), adapted from Ref.~\onlinecite{rakheja2014ambipolar}, are reproduced here for completeness and to drive the discussion that follows.
\begin{subequations}
\begin{equation}
Q_\mathrm{x0e} = C_{g}n_{e}\phi_{t}\log\left(1+\exp{\left(\dfrac{V_\mathrm{gsi}-V_\mathrm{te}}{n_{e}\phi_t}\right)}\right),
\end{equation}
\vspace{-10pt}
\begin{equation}
Q_\mathrm{x0h} = C_{g}n_{h}\phi_{t}\log\left(1+\exp{\left(\dfrac{V_\mathrm{dgi}+V_\mathrm{th}}{n_{h}\phi_{t}}\right)}\right),
\end{equation}
\label{eq:Q_compact}
\end{subequations}
\hspace{-1.5pt}where $C_g$ is the gate-channel capacitance in strong inversion, $V_\mathrm{te}$ ($V_\mathrm{th}$) is the threshold voltage of electron (hole) branch, $n_e$ ($n_h$) is the non-ideality factor of the electron (hole) branch, $V_\mathrm{dgi} = V_\mathrm{dsi}-V_\mathrm{gsi}$. The gate-channel capacitance is $C_g = \epsilon_\mathrm{ox}/CET$, where $\epsilon_\mathrm{ox}$ is the oxide permittivity, and $CET$ is the capacitance equivalent thickness of the oxide. For thick oxides, $CET$ is approximately equal to the physical thickness of the oxide, $t_\mathrm{ox}$. The non-ideality factor incorporates the effect of punch-through ($n_d$) and is given as $n_\mathrm{e(h)} = n_\mathrm{0_{e(h)}} + n_\mathrm{d_{e(h)}}V_\mathrm{dsi}$. Here, $n_0$ is the value of the non-ideality factor when $V_\mathrm{dsi} \rightarrow 0$.

The threshold voltage of the electron and hole branches is given as
\begin{subequations}
\begin{equation}
    V_\mathrm{te} = V_\mathrm{min0} + \Delta V_e - \alpha_e\phi_t FF_e,
\end{equation}
\begin{equation}
    V_\mathrm{th} = V_\mathrm{min0} - \Delta V_h + \alpha_h\phi_t FF_h.
\end{equation}
\end{subequations}
Here, $V_\mathrm{min0}$ corresponds to the ambipolar (minimum-conductivity) point at $V_\mathrm{ds}$ = 0 V, $\Delta V$ approximates the effect of interface traps, and $\alpha_e$ and $\alpha_h$ are empirical fitting parameters. The functions $FF_e$ and $FF_h$ are logistic functions that control the change in the threshold voltage between weak inversion and strong inversion regimes and are given as
\begin{subequations}
\begin{equation}
FF_{e} = \dfrac{1}{1+\exp\left(\dfrac{V_\mathrm{gsi}-\left(V_\mathrm{te}-\dfrac{\alpha_{e}\phi_{t}}{2}\right)}{\alpha_{e}'\phi_{t}}\right)}
\label{eq:FFe},
\end{equation}
\begin{equation}
FF_{h} = \dfrac{1}{1+\exp\left(\dfrac{V_\mathrm{dgi}+\left(V_\mathrm{th}+\dfrac{\alpha_{h}\phi_{t}}{2}\right)}{\alpha_{h}'\phi_{t}}\right)}.
\end{equation}
\label{eq:FF}
\end{subequations}
Here, $\alpha_e'$ and $\alpha_h'$ are introduced as additional fitting parameters to adjust the rate and smoothness of transition of the threshold voltage between its weak inversion and strong inversion limits. Typical values of $\alpha_e$/$\alpha_h$ and $\alpha_e'$/$\alpha_h'$ are in the range of 3-11, depending on the channel length and the type of carriers.

Unlike graphene FETs in which the ambipolar point voltage does not shift with $V_\mathrm{ds}$, \cite{rakheja2013unified} in BP FETs, the ambipolar point voltage shows strong dependence on $V_\mathrm{ds}$. \cite{haratipour2016fundamental, haratipour2015black, haratipour2016ambipolar} Additionally, the thickness of the BP flakes studied in this work is less than 10 nm, which results in a narrow band gap of BP and a significant carrier injection at the drain end. Therefore, the on-off current ratio decreases by over two orders of magnitude as $|V_\mathrm{ds}|$  
increases (see Section~\ref{sec:results} on Results). 
The dependence of $V_\mathrm{min0}$ on $V_\mathrm{ds}$ and the degradation in on-off current ratio with $V_\mathrm{ds}$ can be captured using the following equation:
\begin{eqnarray}
\Delta V_{e(h)} = \Delta_\mathrm{1e(h)} - \Delta_\mathrm{2e(h)}V_\mathrm{dsi} - \Delta_\mathrm{3e(h)}V_\mathrm{dsi}^2,
\label{eq:delta_final}
\end{eqnarray}
where $\Delta_\mathrm{1e(h)}$, $\Delta_\mathrm{2e(h)}$, and $\Delta_\mathrm{3e(h)}$ are extracted from experimental calibration. The threshold voltage model uses 11 parameters, out of which seven parameters: $V_\mathrm{min0}$, $\Delta_\mathrm{1e(h)}$, $\Delta_\mathrm{2e(h)}$, and $\Delta_\mathrm{3e(h)}$, can be extracted from the transfer curves by measuring the change in the threshold voltage with $V_\mathrm{dsi}$ (see Appendix A for details.) The remainder four parameters: $\alpha_{e(h)}$,  $\alpha_{e(h)}'$ are tweaked in the range around 3-11 to match the transition between the on and off characteristics of the transistor, as well as to adjust the smoothness of the device transconductance. The validation of the charge model and gate capacitance against numerical data is discussed in Appendix C.

\vspace{-10pt}
\subsection{Contact resistance} \label{sec:contact}
\vspace{-10pt}
The parasitic resistances associated with source and drain contacts in BP FETs are bias-dependent, nonlinear resistances. This is because at the interface between the contact metal and the BP channel, a Schottky barrier is formed, which shows a strong bias dependence.~\cite{chang2014mobility, chang2017germanium, du2016transport}

The current, $I_\mathrm{SB}$, through a Schottky barrier (SB) for a given voltage drop $V_\mathrm{SB}$ across it is given as~\cite{xu2016contacts, chang2014mobility}
\begin{equation}
I_\mathrm{SB} = A_\mathrm{jun}A_{R}T^2\exp{\left(-\dfrac{\phi_B}{\phi_t}\right)}\left[\exp\left(\dfrac{V_\mathrm{SB}}{\eta_\mathrm{SB} \phi_t}\right)-1\right],
\label{eq:I_tunn}
\end{equation}
where $A_\mathrm{jun}$ is the effective junction area of the SB contact, $A_{R}$ is the Richardson constant, and $\phi_{B}$ is the SB height. 
For purely thermionic emission at the SB contact, the non-ideality factor $\eta_\mathrm{SB}$ = 1. For other types of current conduction, such as thermally assisted tunneling and Fowler–Nordheim tunneling, $\eta_\mathrm{SB}>1$. Other factors that may lead to $\eta_\mathrm{SB}>1$ include bias dependence and image force lowering of the SB height, generation and recombination of carriers at the SB contact, and in-homogeneity of the junction.~\cite{xu2016contacts}
Equation~(\ref{eq:I_tunn}) shows that the SB current is exponentially dependent on the SB height and the voltage drop across the barrier. The analytic model of contact resistance developed here captures these key aspects of current conduction through an SB contact. Moreover, in back-gated BP FETs we study, the region underneath the contacts is intrinsically p-doped. The application of a large negative gate voltage increases the hole doping under the contacts, which leads to a narrower barrier for hole injection and reduces the contact resistance corresponding to the hole branch.

\begin{figure}[H]
 \centering
\includegraphics [width=3.5in]{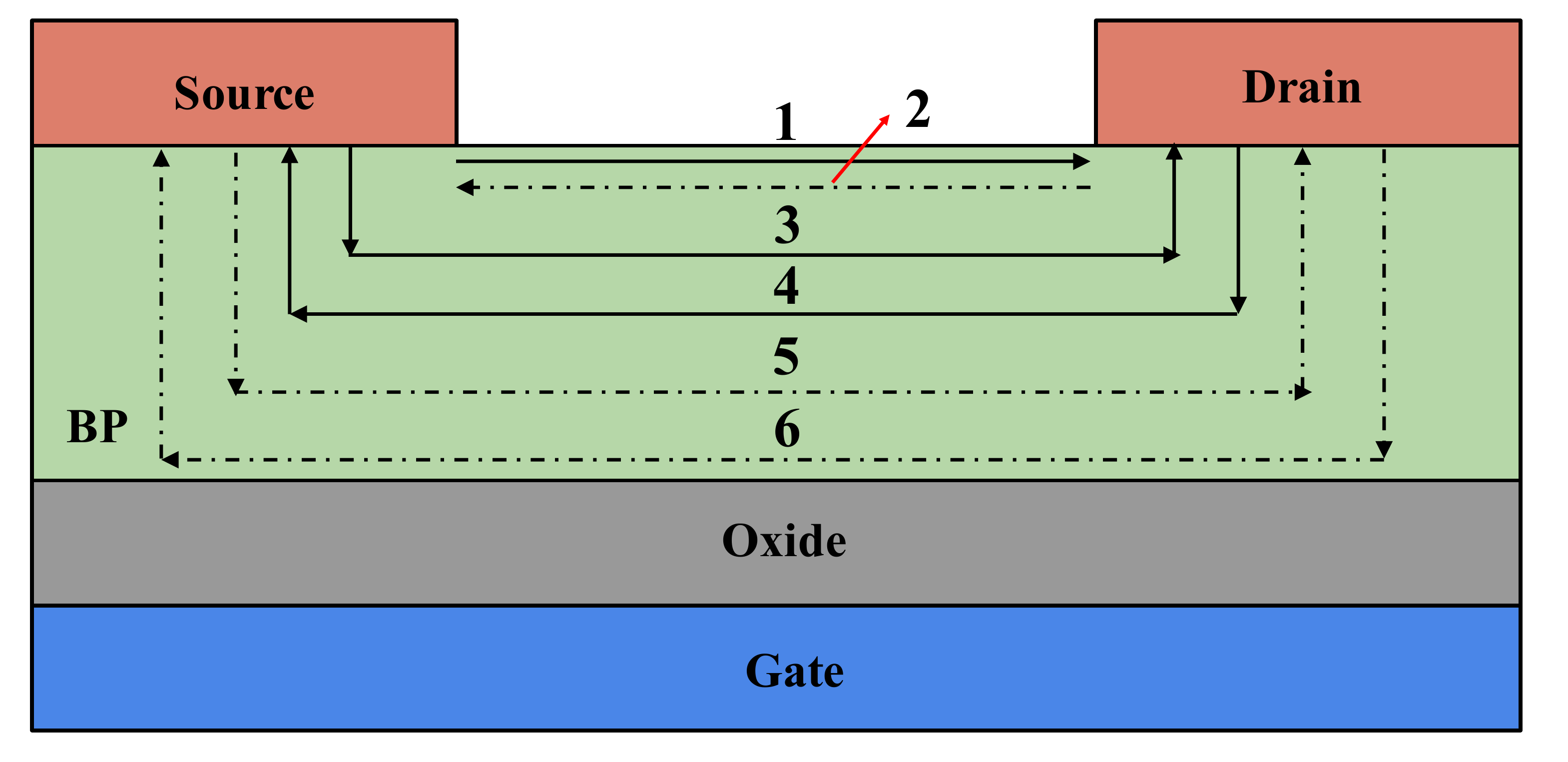}
\vspace{-5pt}
\caption{Possible carrier transport paths in the channel. Solid and dashed lines represent the flow of electrons and holes, respectively, through the channel. Note that the figure is not drawn to scale.}
\label{fig:path}
\vspace{-5pt}
\end{figure}

To fully understand the effect of SB contacts on current conduction, we focus on the various paths in Fig.~\ref{fig:path} that the carriers, once injected from the contacts, can take through the BP channel. In this figure, solid and dashed lines represent the flow of electrons and holes, respectively, through the channel. Note that in the model, we call electrical source as the terminal with lower voltage and electrical drain the terminal with higher voltage.

In Fig.~\ref{fig:path}, path 1 for electrons and path 2 for holes are transparent for all values of $V_\mathrm{gs}$ when $V_\mathrm{ds}>$0. For $V_\mathrm{ds}>$0, $V_\mathrm{gs}>$0, and $V_\mathrm{ds}<V_\mathrm{gs}$, paths labeled as $3$ and $4$ are unavailable for electron conduction. Likewise for hole, paths labeled as $5$ and $6$ are cut-off. Hence, we can conclude that for $V_\mathrm{ds}<V_\mathrm{gs}$, the only way carriers can be transported between the contacts is through the paths labeled as 1 and 2 in Fig.~\ref{fig:path}. Results of this discussion are summarized in Table~\ref{tab:table2}.
\begin{table}[h!]
\vspace{-5pt}
\caption{\label{tab:table2} Possibility of different carrier transport paths in a Schottky barrier back- gated MOSFET for $V_\mathrm{ds}>0$, $V_\mathrm{gs}>0$ and $V_\mathrm{ds}<V_\mathrm{gs}$.}
\begin{tabular}{|p{1.5cm}|p{1.5cm}|p{1.5cm}|p{1.5cm}|p{2cm}|}
 \hline
\textbf{Carrier} & \textbf{From} & \textbf{To} & \textbf{Path} & \textbf{Possibility}\\\hline\hline
Electron & Source & Drain & 1 & \cmark \\\hline
Electron & Source & Drain & 3 & \xmark\\\hline
Electron & Drain & Source & 4 & \xmark\\\hline
Hole & Drain & Source & 2 & \cmark\\\hline
Hole & Source & Drain & 5 & \xmark\\\hline
Hole & Drain & Source & 6 & \xmark\\\hline
\end{tabular}
\vspace{-5pt}
\end{table}

Next, we consider the case when $V_\mathrm{gs}<V_\mathrm{ds}$, $V_\mathrm{ds}>0$, and $V_\mathrm{gs}>0$. In this case, in addition to paths labeled as 1 and 2 in Fig.~\ref{fig:path}, there exist path $3$ for electron conduction and path $6$ for hole conduction. Results are summarized in Table~\ref{tab:table3}. 
\begin{table}[h!]
\vspace{-5pt}
\caption{\label{tab:table3} Possibility of different carrier transport paths in a Schottky barrier back- gated MOSFET for $V_\mathrm{ds}>0$, $V_\mathrm{gs}>0$, and $V_\mathrm{ds}>V_\mathrm{gs}$.}
\begin{tabular}{|p{1.5cm}|p{1.5cm}|p{1.5cm}|p{1.5cm}|p{2cm}|}
 \hline
\textbf{Carrier} & \textbf{From} & \textbf{To} & \textbf{Path} & \textbf{Possibility}\\\hline\hline
Electron & Source & Drain & 1 & \cmark \\\hline
Electron & Source & Drain & 3 & \cmark\\\hline
Electron & Drain & Source & 4 & \xmark\\\hline
Hole & Drain & Source & 2 & \cmark\\\hline
Hole & Source & Drain & 5 & \xmark\\\hline
Hole & Drain & Source & 6 & \cmark\\\hline
\end{tabular}
\vspace{-5pt}
\end{table}

An appropriate bias-dependent model of the SB contact resistance in BP FETs must comprehend the distinct behavior for $V_\mathrm{gs}>V_\mathrm{ds}$ and $V_\mathrm{gs}<V_\mathrm{ds}$ regimes of transport when $V_\mathrm{ds}>0$ and $V_\mathrm{gs}>0$. When $V_\mathrm{ds}>0$ and $V_\mathrm{gs}<0$ the possible carrier transport paths are the same as the results of the Table~\ref{tab:table2}. Note that the bias dependence of SB contact resistance is exacerbated for hole conduction. This is because of the use of Ti as the metal contact in the experimental BP FET devices examined here. Compared to metals, such as Pd or Ni, Ti has a much lower work-function, which gives rise to a larger SB height and, therefore, large contact resistance values.~\cite{das2014ambipolar, haratipour2017high} Also, the effect of contact resistance corresponding to hole conduction becomes especially significant in short channel devices in which the contact resistance could easily dominate the total drain-source resistance, limiting the maximum available current through the device.~\cite{valletta2011contact} Here, we assume that $R_\mathrm{elec}$ is a bias-independent, linear resistance. This is justified because of the p-type background doping in the devices under study.

The hole branch contact resistance assuming both ohmic resistance and the formation of the SB barrier at the metal/BP interface is given as
\begin{multline}
 R_\mathrm{hole} = \underbrace{R_\mathrm{ohmic,1}FF_{R} + R_\mathrm{ohmic,2}\left(1-FF_{R}\right)}_\text{Ohmic contact resistance}+    \underbrace{\left(R_\mathrm{01}FF_{R} + R_\mathrm{02}\left(1-FF_{R}\right)\right)\exp{\left(aV_\mathrm{gsi}\right)}}_\text{Schottky barrier resistance},
 \label{multline:R}
\end{multline}
where $a = a_{1}FF_{R} + a_{2}\left(1-FF_{R}\right).$

In the above set of equations, we use the logistic function $FF_R$ to model the drain-bias dependence of various parameters. This function is similar to the $FF_e$/$FF_h$ logistic function used in Eq. (\ref{eq:FF}) and is given as
\begin{eqnarray}
FF_R = \dfrac{1}{1 + \exp{\left(\dfrac{-V_\mathrm{dsi} + V_0}{\gamma}\right)}},
\label{eq:FF_R}
\end{eqnarray}
where
$\gamma$ is used to adjust the sharpness of transition between two different bias regions, and $V_{0}$ is the drain-gate voltage at which additional paths for current conduction between the contacts are introduced (see Fig.~\ref{fig:path} and the discussion.) This voltage is given as
\begin{eqnarray}
V_{0} = V_\mathrm{01}FF_{h} + V_\mathrm{02}\left(1-FF_{h}\right),
\label{eq:V0}
\end{eqnarray}
where $V_\mathrm{01}$ and $V_\mathrm{02}$ are fitting parameters, and $FF_{h}$ is given in Eq.~(\ref{eq:FF}b).
The model for $R_\mathrm{hole}$ described here captures the exponential dependence on $V_\mathrm{gsi}$ as expected for SB contacts. Moreover, the model can also explain the drain-bias dependence of the contact resistance, following the discussion pertinent to Fig.~\ref{fig:path}. The contact resistance model introduces 10 parameters: $R_\mathrm{elec}$, $R_\mathrm{ohmic,1}$, $R_\mathrm{ohmic,2}$, $R_{01}$, $R_{02}$, $V_{01}$, $V_{02}$, $\gamma$, $a_1$, $a_2$. The methodology to
extract the contact resistance parameters from experimental
data is presented in Appendix A.

\vspace{-10pt}
\section{Effect of temperature}\label{sec:temp}
\vspace{-10pt}
The main model parameters that are affected by temperature include the mobility, saturation velocity, non-ideality factor, and threshold voltage. Below we discuss the models to capture the temperature dependence of the various parameters.

\noindent
\paragraph{Mobility:}
Experimentally measured mobility of carriers in 2D materials is generally much lower than that predicted theoretically based on phonon-dominated collision. Mobility degradation in 2D materials results from defects and charged impurities at room temperature.\cite{ong2014anisotropic, jariwala2014emerging, radisavljevic2013mobility,ovchinnikov2014electrical}
Previous published works indicate that the mobility of carriers in BP follows a power law relationship with respect to temperature ($T$). That is, $\mu \propto T^{-\xi_\mu}$ for $T>$100 K \cite{ong2014anisotropic, trushkov2017phonon, li2014black, xia2014rediscovering} with $\xi_\mu$ typically in the range of -0.4 to 1.2 based on the type and concentration of carriers, BP crystal orientation, and the dielectric environment of the sample.\cite{ong2014anisotropic} 
In this paper, we use a constant carrier mobility model with temperature dependence given as
 \begin{equation}
 \mu_{e/h} = \mu_{298,e/h}\left(\frac{298}{T}\right)^{\xi_{\mu,{e/h}}}.
 \label{eq:mu}
 \end{equation}
Here, $\mu_{298, e/h}$ is the value of the mobility of carriers at 298 K. 
As a result of the weakened polarization charge screening for oxides with a high dielectric constant, such as HfO$_2$, we expect the temperature dependence of mobility in BP devices under study to be weak.~\cite{ong2014anisotropic} As such, $\xi_{\mu,e/h}$ is expected to be in the range of 0.01 to 0.3.

The effect of carrier concentration on mobility, as demonstrated in recent experimental work,\cite{haratipour2018mobility} is studied in Appendix B. However, for validation of the model against experimental data, we consider a constant carrier mobility model as in the equation above. This allows us to restrict the number of model parameters without compromising the quality of the model fits while also providing reasonable estimates of extracted parameters (see Sec.~\ref{sec:results} on results.)

\noindent
\paragraph{Saturation velocity:}
The saturation velocity of carriers in BP is expected to decrease with an increase in temperature due to enhanced phonon-dominated scatterings.~\cite{chen2018large}
Here, we model $v_\mathrm{x0}$ of both electrons and holes using a linear equation in the range of 180 K to 298 K. This model is similar to that used previously.~\cite{chen2018large}
\begin{equation}
v_{\mathrm{x0},e/h} = v_{\mathrm{x0,298},e/h} - \xi_{v,e/h}\left(T - 298\right),
\label{eq:vx0}
\end{equation}
where $v_{\mathrm{x0,298},e/h}$ is the saturation velocity of carriers at 298 K, and $\xi_{v,e/h}$ is the temperature coefficient of saturation velocity. Typical value of $\xi_v$ is in the range of $50$ to $100$ {m/sK}.

\noindent
\paragraph{\label{sec:n}Non-ideality factor:}
At low $V_\mathrm{ds}$, sub-threshold slope is modeled as $SS = 2.3n\phi_t$ where $n$ is the
non-ideality factor 
defined previously as $n = n_0 + n_d V_\mathrm{dsi}$.
Experimental results in Section~\ref{sec:results} indicate that for low-$V_\mathrm{ds}$, $SS$ is linearly proportional to temperature,~\cite{haratipour2016fundamental} implying that $n_0$ is independent of temperature.
At high $V_\mathrm{ds}$, tunneling current through the drain contact dominates, and the dependence of $SS$ on temperature becomes sub-linear. This behavior can be reproduced by considering a linear temperature variation in the punch-through factor, $n_d$:
\begin{equation}
n_{d,e/h} = n_{\mathrm{d,298},e/h} - \xi_\mathrm{n_{d,{e/h}}}\left(T - 298\right),
\label{eq:nd}
\end{equation}
where $n_{\mathrm{d,298},e/h}$ is the value of punch-through factor at 298 K, and $\xi_\mathrm{n_{d,{e/h}}}$ captures the temperature sensitivity of $n_d$. Typical values of $\xi_\mathrm{n_{d,{e/h}}}$ are in the range of 0.009 to 0.025 {1/VK}.

\noindent
\paragraph{Threshold voltage:}
The temperature dependence of threshold voltage is introduced in the parameter $\Delta_{1,e/h}$ used in Eq. (\ref{eq:delta_final}) according to
\begin{equation}
\Delta_{1,e/h} = \Delta_{\mathrm{1,298},e/h}-\xi_\mathrm{\Delta_{1,e/h}}\left(T-298\right),
\label{eq:Delta_1}
\end{equation}
$\Delta_{\mathrm{1,298},e/h}$ is the value of the parameter at $T =$ 298 K, and $\xi_\mathrm{\Delta_{1,e/h}}$ captures the temperature dependence of $\Delta_{1,e/h}$ and it is on the order of $10^{-3}$. The linear variation of $\Delta_{1,e/h}$ with temperature reproduces experimental and numerical results accurately as discussed in Section~\ref{sec:results}.

\vspace{-10pt}
\section{Results} \label{sec:results}
\vspace{-10pt}
The analytic I-V model developed here has 39 parameters, out of which 20 (11) parameters correspond to hole (electron) conduction, and 8 parameters (4 parameters for each carrier type) model the temperature sensitivity of current conduction. These models can be obtained through a systematic experimental validation methodology as explained in Appendix A.

There are four datasets available from experimental measurements and numerical simulation using the TCAD simulation tool Sentaurus from Synopsys.\cite{guide2016version}
The first two datasets correspond to the back-gated BP FETs with Schottky source/drain contacts which are fabricated at the University of Minnesota. In these two datasets, the BP flakes are exfoliated from the bulk
crystal and transferred onto the local back gates.
The gate dielectric is HfO$_2$ with a thickness of 15 nm, and the gate metal is Ti(10 nm)/Pd(40 nm). The source and drain contacts are Ti (10 nm)/Au (90 nm). The device is passivated using 20-30 nm of Al$_2$O$_3$ to protect BP from atmospheric degradation. Additional fabrication details are given in Ref.~\onlinecite{haratipour2016fundamental}. 
The third and fourth dataset are obtained numerically by simulating top and back-gated BP FETs with Schottky source/drain contacts to show model validity for both FETs structures. 
Model parameters extracted for all datasets are listed in Tables~\ref{tab:table_t} and ~\ref{tab:table_dataset3}.

For all datasets that are discussed in this section, we refer to the terminal that is grounded as the source terminal ($V_s =$ 0 V) and the drain terminal is biased at negative voltages ($V_d<$0 V). Note that this terminology of labeling the source/drain contacts is different from that followed conventionally, but it does not impact the interpretation of the model and the results. The implementation of the model in Verilog-A, SPICE circuit simulations, and Gummel symmetry test are presented in Appendix D.

\begin{figure*}
\centering
\vspace{-10pt}
{\textbf{Dataset 1: $L = 1$ $\mu$m, $W = 7.32$ $\mu$m, $t_\mathrm{BP} = 7.3$ nm, $t_\mathrm{ox} = 15$ nm, with rotational angle $42^{\circ}$, and $T = 298$ K}}
\vspace{-10pt}
\centering\includegraphics[width=16cm]{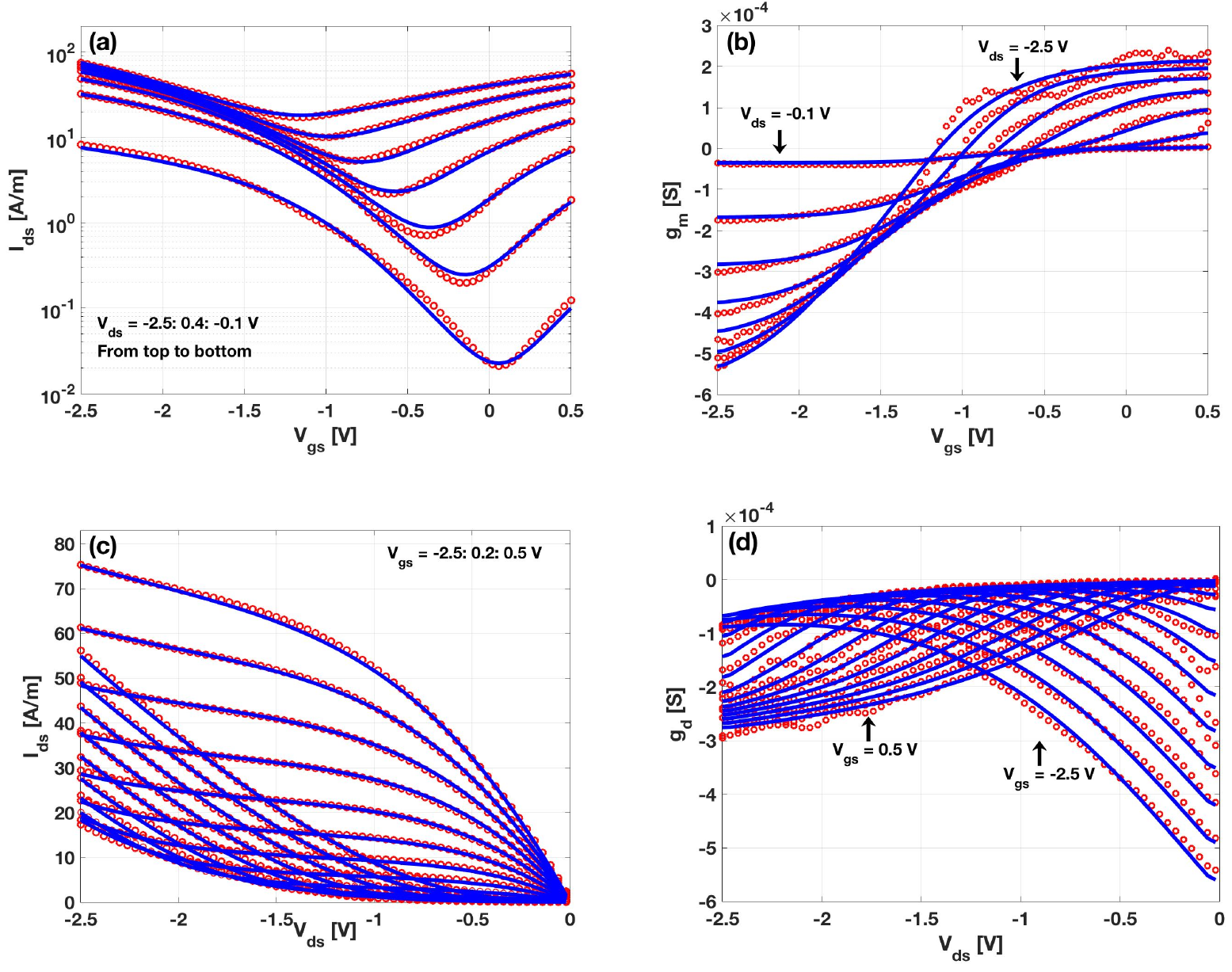}
\vspace{-5pt}
  \caption{\label{fig:IVdg6} Model fit to the experimental dataset 1 at room temperature: (a) transfer characteristics ($I_\mathrm{ds}$-$V_\mathrm{gs}$), (b) transconductance ($g_m = \partial I_\mathrm{ds}/\partial V_\mathrm{gs}$), (c) output characteristics ($I_\mathrm{ds}$-$V_\mathrm{ds}$), and (d) output conductance ($g_d = \partial I_\mathrm{ds}/\partial V_\mathrm{ds}$). Solid lines are model fits, while symbols correspond to experimental data.}
\vspace{-10pt}
\end{figure*}

\begin{figure*}
\vspace{-10pt}
{Dataset 1: $L = 1$ $\mu$m, $W = 7.32$ $\mu$m, $t_\mathrm{BP} = 7.3$ nm, $t_\mathrm{ox} = 15$ nm, with rotational angle $42^{\circ}$ at different temperatures.}
\vspace{-10pt}  
\centering\includegraphics[width=16cm]{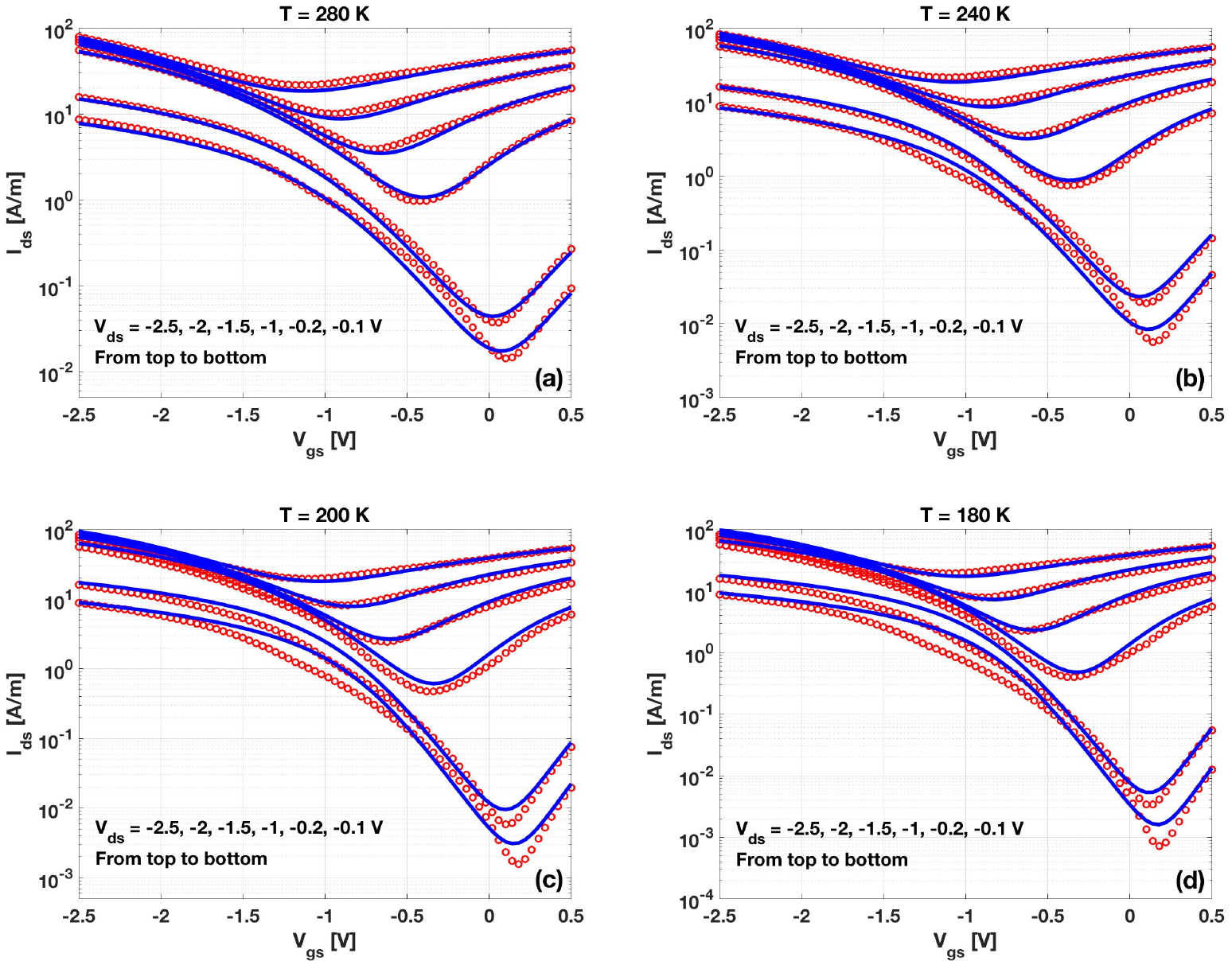}
\vspace{-1pt}
\caption{\label{fig:temp6} Model fit to the experimental dataset 1 at $T =$ (a) 280 K, (b) 240 K, (c) 200 K, and (d) 180 K for $V_\mathrm{ds}:$ -0.1, -0.2, -1, -1.5, -2, and -2.5 V. Solid lines are model fits, while symbols correspond to experimental data.}
\vspace{-10pt}
\end{figure*}

\begin{figure}[h!]
\vspace{-5pt}
{Dataset 1: $L = 1$ $\mu$m, $W = 7.32$ $\mu$m, $t_\mathrm{BP} = 7.3$ nm, $t_\mathrm{ox} = 15$ nm, with rotational angle $42^{\circ}$}
\vspace{-5pt}
\includegraphics [width=8.6cm, height=6.5cm]{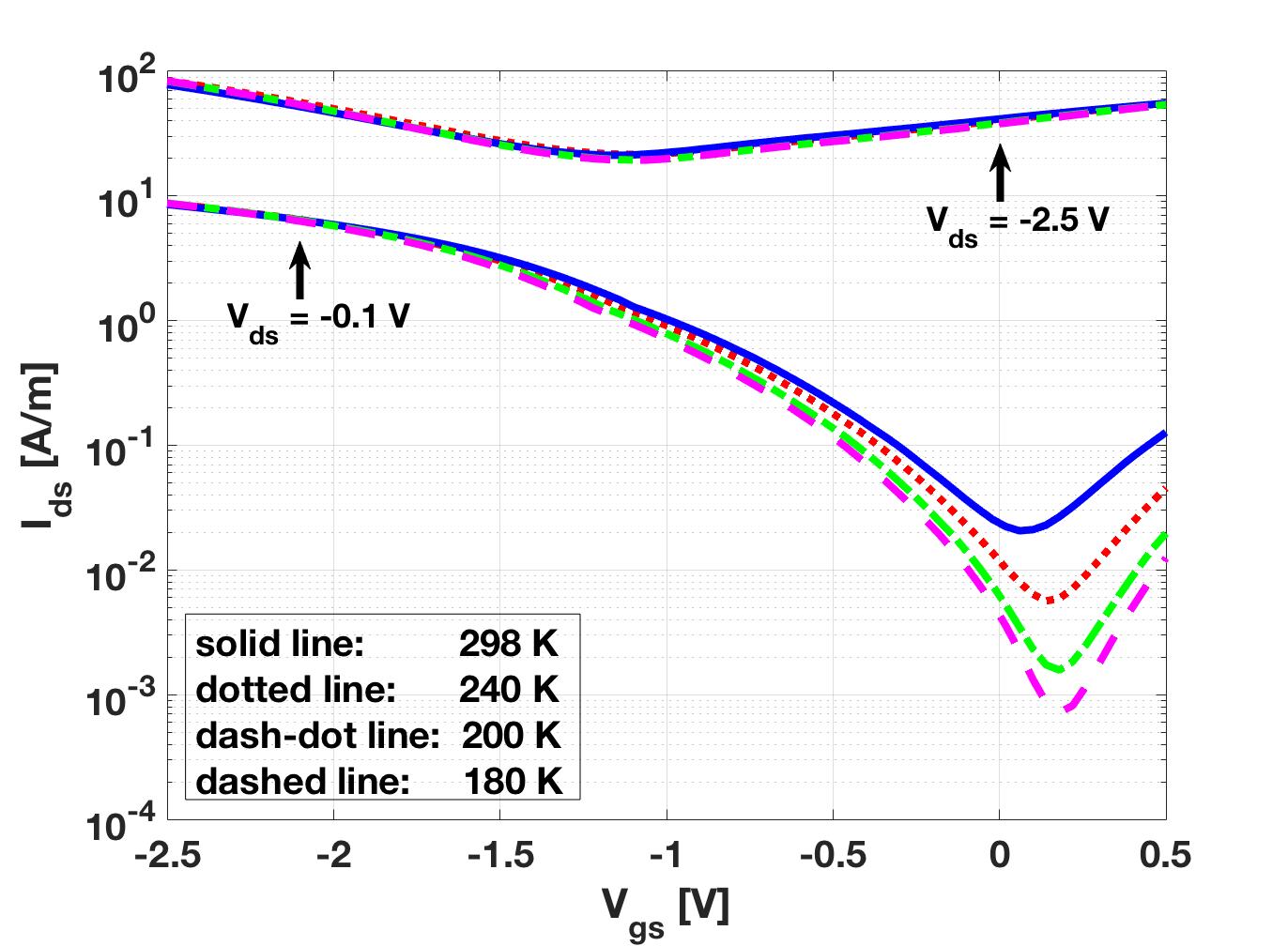}
\vspace{-5pt}
\caption{\label{fig:diff_temp} Experimental data for transfer characteristics ($I_\mathrm{ds}$-$V_\mathrm{gs}$) of dataset1 for $V_\mathrm{ds} = -0.1, -2.5$ V at different temperatures. Solid line, dotted line, dash-dot line, and dashed line correspond to the $T = 298$ K, $240$ K, $200$ K, and $140$ K, respectively.}
\vspace{-5pt}
\end{figure}

\begin{figure}[h!]
 \vspace{3pt}
{Dataset 2: $L = 0.3$ $\mu$m, $W = 3.16$ $\mu$m, $t_\mathrm{BP} = 8.1$ nm, $t_\mathrm{ox} = 15$ nm, with rotational angle of $40^{\circ}$.}
\vspace{-5pt}
\includegraphics [width=8.5cm, height=6cm]{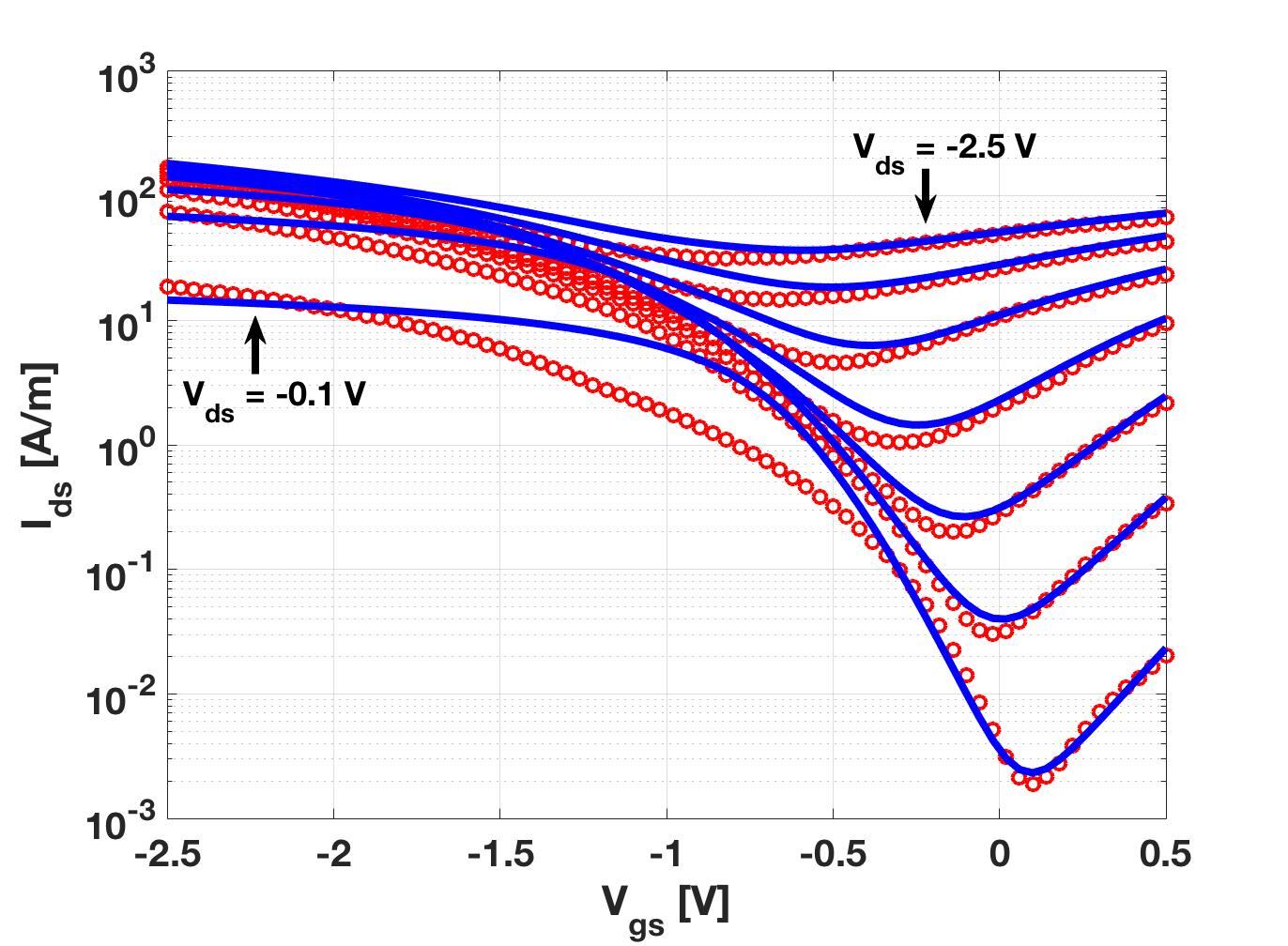}
\vspace{-5pt}
\caption{\label{fig:const_R}Transfer characteristics ($I_\mathrm{ds}$-$V_\mathrm{gs}$) of dataset 2 with constant $R_\mathrm{hole}$ and $R_\mathrm{elec}$ at room temperature. Solid lines are model fits, while symbols correspond to experimental data.}
\vspace{-5pt}
\end{figure}

\begin{figure*}
\vspace{-10pt}
{Dataset 2: $L = 0.3$ $\mu$m, $W = 3.16$ $\mu$m, $t_\mathrm{BP} = 8.1$ nm, $t_\mathrm{ox} = 15$ nm, with rotational angle of $40^{\circ}$, and $T = 298$ K.}
\vspace{-10pt}
\centering\includegraphics[width=16cm]{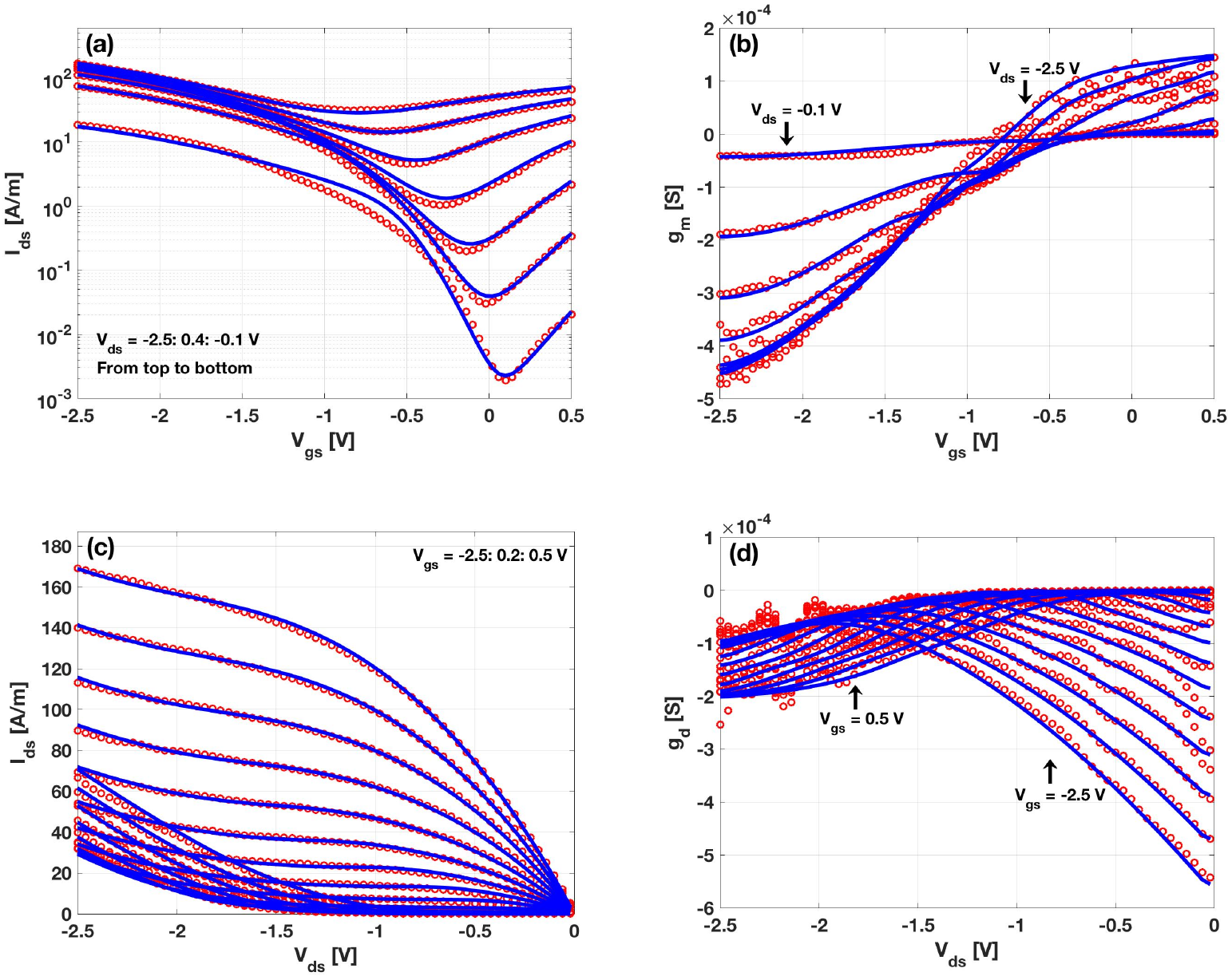}
  \caption{\label{fig:IVdg7} Model fit to the experimental dataset 2 at room temperature: (a) transfer characteristics ($I_\mathrm{ds}$-$V_\mathrm{gs}$), (b) transconductance ($g_m = \partial I_\mathrm{ds}/\partial V_\mathrm{gs}$), (c) output characteristics ($I_\mathrm{ds}$-$V_\mathrm{ds}$), and (d) output conductance ($g_d = \partial I_\mathrm{ds}/\partial V_\mathrm{ds}$). Solid lines are model fits, while symbols correspond to experimental data.}
\vspace{-10pt}
\end{figure*}

\begin{figure*}
\vspace{-10pt}
{Dataset 2: $L = 0.3$ $\mu$m, $W = 3.16$ $\mu$m, $t_\mathrm{BP} = 8.1$ nm, $t_\mathrm{ox} = 15$ nm, with rotational angle of $40^{\circ}$.}
\vspace{-10pt}
    \centering\includegraphics[width=16cm]{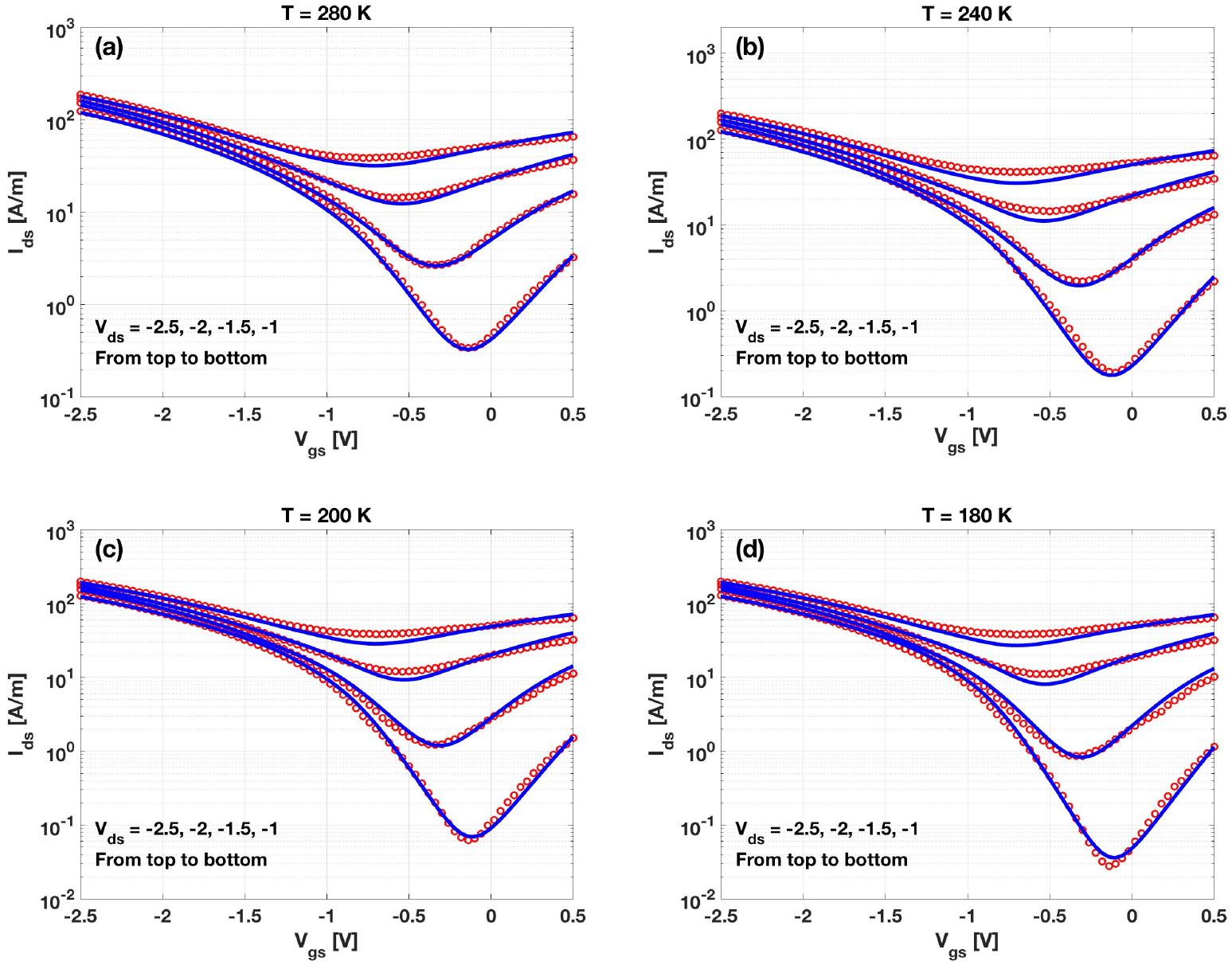}
  \caption{\label{fig:temp7} Model fit to the experimental dataset 2 at $T =$ (a) 280 K, (b) 240 K, (c) 200 K, and (d) 180 K for $V_\mathrm{ds}:$ -1, -1.5, -2, and -2.5 V. Solid lines are model fits, while symbols correspond to experimental data.}
\vspace{-10pt}
\end{figure*}

\begin{figure*}
\vspace{-10pt}
{Dataset 3: $L = 0.3$ $\mu$m, $W = 3.16$ $\mu$m, $t_\mathrm{BP} = 8.1$ nm, $t_\mathrm{ox, back} = 15$ nm, $t_\mathrm{ox,top} = 10$ nm, $L_\mathrm{cont} = 1.16$ $\mu$m, and $T = 298$ K}
\vspace{-10pt} 
    \centering\includegraphics[width=16cm]{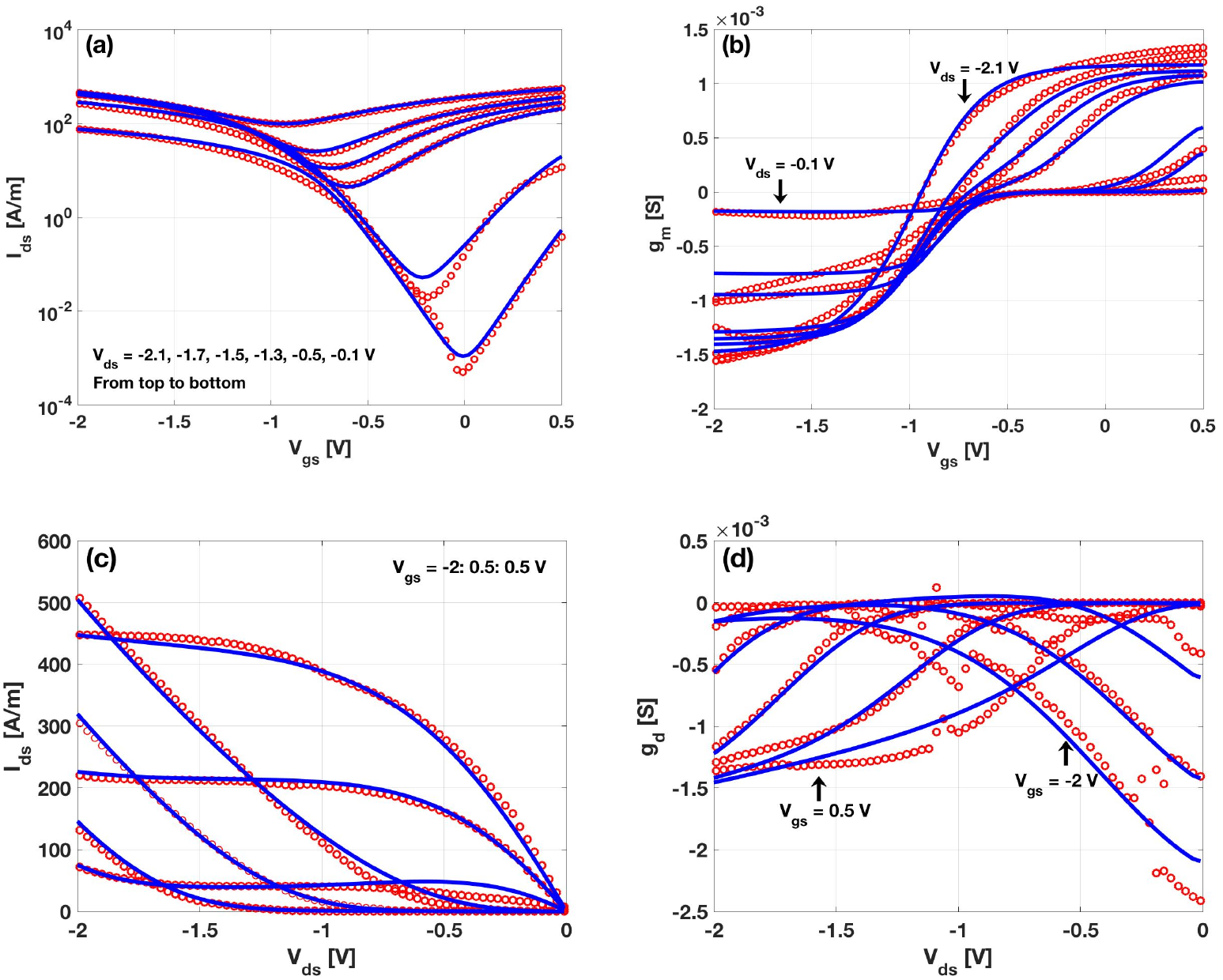}
  \caption{\label{fig:IVdg_topG} Model fit to the numerical dataset 3 at room temperature: (a) transfer characteristics ($I_\mathrm{ds}$-$V_\mathrm{gs}$), (b) transconductance ($g_m = \partial I_\mathrm{ds}/\partial V_\mathrm{gs}$), (c) output characteristics ($I_\mathrm{ds}$-$V_\mathrm{ds}$), and (d) output conductance ($g_d = \partial I_\mathrm{ds}/\partial V_\mathrm{ds}$). Solid lines are model fits, while symbols correspond to numerical data.}
\vspace{-10pt}
\end{figure*}

\begin{figure*}
\vspace{-10pt}
{Dataset 3: $L = 0.3$ $\mu$m, $W = 3.16$ $\mu$m, $t_\mathrm{BP} = 8.1$ nm, $t_\mathrm{ox, back} = 15$ nm, $t_\mathrm{ox,top} = 10$ nm, and $L_\mathrm{cont} = 1.16$ $\mu$m.}
\vspace{-10pt}
    \centering\includegraphics[width=16cm]{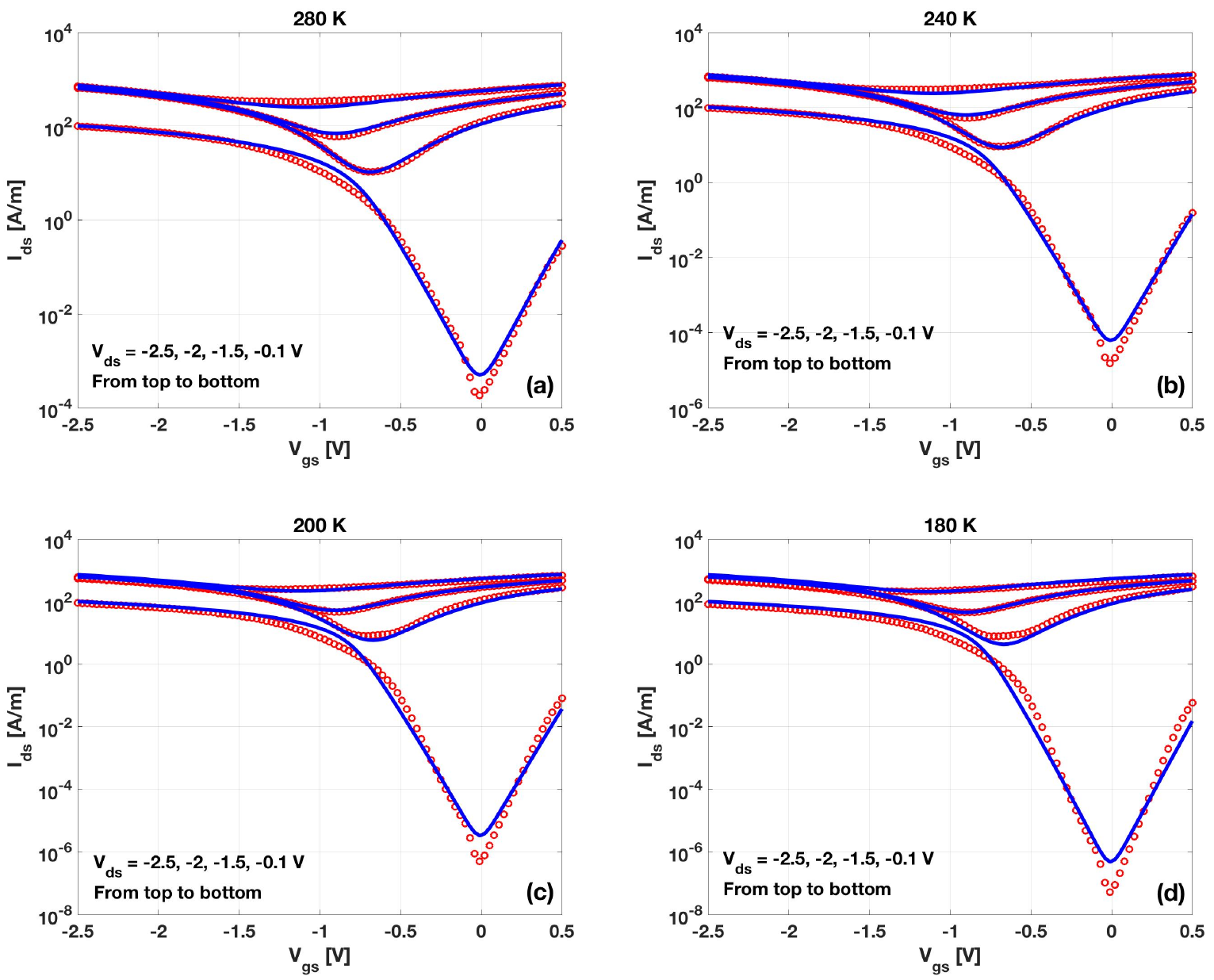}
  \caption{\label{fig:temp_topG} Model fit to the numerical dataset 3 at $T =$ (a) 280 K, (b) 240 K, (c) 200 K, and (d) 180 K for $V_\mathrm{ds}:$ -0.1, -1.5, -2, and -2.5 V. Solid lines are model fits, while symbols correspond to numerical data.}
\vspace{-10pt}
\end{figure*}

 \begin{table}

\caption{\label{tab:table_t} Model parameters for datasets 1 and 2 (experimental).}
\begin{ruledtabular}
{\scriptsize
\begin{tabular}{lcccc}
 &\multicolumn{2}{c}{\textbf{Dataset 1}}&\multicolumn{2}{c}{\textbf{Dataset 2}}\\
 \textbf{Model parameter} & \textbf{Electron} & \textbf{Hole} & \textbf{Electron}
& \textbf{Hole}\\ \hline
Carrier mobility at 298 K, $\mu_\mathrm{298}$ ({cm$^2$/Vs}) &19 & 49 & 40 & 60 \\\hline
Temperature dependence for mobility, $\xi_{\mu}$ (unit-less) & 0.01 & 0.17 & 0.2 & 0.3 \\\hline
Carrier saturation velocity at 298 K, $v_\mathrm{x0,298}$ ({m/s}) & $0.5\times10^{4}$ & $1.5\times10^{4}$ & $1.5\times10^{4}$ & $2\times10^{4}$\\\hline
Temperature dependence for saturation velocity, $\xi_v$ ({m/sK})& 50 & 50 & 50 & 50 \\\hline
Non-ideality factor, $n_0$ (unit-less)& 7 & 7.5 & 5.4 & 2.5 \\\hline
Punch-through factor at 298 K, $n_\mathrm{d,298}$ ({1/V}) & 0.12 & 3.42 & 3 & 3.5 \\\hline
Temperature dependence for punch-through factor, $\xi_\mathrm{nd}$ ({1/VK}) & 0.025 & 0.02 & 0.009 & 0.013 \\\hline
Shift in threshold voltage for charge trapping at 298 K, $\Delta_\mathrm{1,298}$ (V) & 0.67 & 1.02 & 1.02 & 0.71 \\\hline
Temperature dependence for $\Delta_1$, $\xi_\mathrm{\Delta_1}$ (V/K) & $1.2\times10^{-3}$ & $-1.64\times10^{-3}$ & $0.44\times10^{-3}$ & 0\\\hline 
Shift in ambipolar point at high $|V_\mathrm{ds}|$, $\Delta_2$ (1/V) & 0.14 & 0.52 & 1.05 & 0.79\\\hline
Threshold voltage adjustment parameter at high $|V_\mathrm{ds}|$, $\Delta_3$  & 0.002 & 0.083 & 0.4 & 0.28\\\hline
Ambipolar point at $V_\mathrm{ds} = 0$, $V_\mathrm{min0}$ (V) & 0.162 & 0.162 & 0.181 & 0.181\\\hline
Adjustment the smoothness of $V_\mathrm{te(h)}$ transition, $\alpha'$ (unit-less) & 9 & 9 & 4 &  6\\\hline
Shift in the $V_\mathrm{te(h)}$ in sub-threshold and strong inversion, $\alpha$ (unit-less)& 3 & 3 & 2 & 6 \\\hline
Empirical parameter for $F_\mathrm{sat}$, $\beta$ (unit-less)& 1.5 & 1.5 & 1.5 & 2 \\\hline
Contact resistance  at 298 K, $R_\mathrm{elec(hole)}$ ({$\Omega$-m})& $10^{-8}$ & $10^{-8}$ & $65\times10^{-4}$ & ---\\\hline
Contact ohmic resistance, $R_\mathrm{ohmic1(2)}$ ({$\Omega$-m}) & --- & ---& ---& $3\times10^{-4}$($14\times10^{-4}$)\\\hline
Schottky barrier resistance coefficient, $R_\mathrm{01(2)}$ ($\Omega$-m) & --- & ---& ---& 0.105(0.014)\\\hline
Gate voltage dependence for Schottky barrier resistance, $a_\mathrm{1(2)}$ (\SI{}{1/V})& --- & ---& ---& 2.053(1.39)\\\hline
Transition point between two different transport paths, $V_\mathrm{01(2)}$ (\SI{}{V})& ---& ---& ---& 0(0.452)\\\hline
Adjustment the smoothness of $R_\mathrm{hole}$, $\gamma$ (\SI{}{V}) & ---& ---& ---& 0.67\\
\end{tabular}}

\end{ruledtabular}

\end{table}

\begin{table}
\caption{\label{tab:table_dataset3} Model parameters for datasets 3 and 4 (numerical).}
\begin{ruledtabular}
{\scriptsize
\begin{tabular}{lcccc}
&\multicolumn{2}{c}{\textbf{Dataset 3}}&\multicolumn{2}{c}{\textbf{Dataset 4}}\\
 \textbf{Model parameter} & \textbf{Electron} & \textbf{Hole} & \textbf{Electron}
& \textbf{Hole} \\ \hline
Carrier mobility at 298 K, $\mu_\mathrm{298}$ (\SI{}{cm^2/Vs}) &110 & 145 &110 & 145 \\\hline
Temperature dependence for mobility, $\xi_{\mu}$ (unit-less) & 0.05 & 0.15 & ---&--- \\\hline
Carrier saturation velocity at 298 K, $v_\mathrm{x0,298}$ (\SI{}{m/s}) & $2.9\times10^{4}$ & $3.9\times10^{4}$ & $2.9\times10^{4}$ & $3.9\times10^{4}$ \\\hline
Temperature dependence for saturation velocity, $\xi_v$ (\SI{}{m/sK})& 50 & 50 & ---& ---\\\hline
Non-ideality factor, $n_0$ (unit-less)& 2.5 & 2.5 & 6 & 3\\\hline
Punch-through factor at 298 K, $n_\mathrm{d,298}$ (\SI{}{1/V}) &2.5 & 1 & 3 & 1\\\hline
Temperature dependence for punch-through factor, $\xi_\mathrm{nd}$ (\SI{}{1/VK}) & 0.0152 & 0.016 & ---& --- \\\hline
Shift in threshold voltage for charge trapping at 298 K, $\Delta_\mathrm{1,298}$ (\SI{}{V}) & 0.69 & 0.73 & 2 & 1\\\hline
Temperature dependence for $\Delta_1$, $\xi_\mathrm{\Delta_1}$ (\SI{}{V/K}) & $0.76\times10^{-3}$ & $0.5\times10^{-3}$ & --- & ---\\\hline 
Shift in ambipolar point at high $|V_\mathrm{ds}|$, $\Delta_2$ (\SI{}{1/V}) & 0.62 & 0.37 & 0.59 & 0.6\\\hline
Threshold voltage adjustment parameter at high $|V_\mathrm{ds}|$, $\Delta_3$ (1/V$^2$) & 0.2 & 0.105 & 0.1 & 0.5\\\hline
Ambipolar point at $V_\mathrm{ds} = 0$, $V_\mathrm{min0}$ (V) & 0.02 & 0.02 & 0.17 & 0.17\\\hline
Adjustment the smoothness of $V_\mathrm{te(h)}$ transition, $\alpha'$ (unit-less) & 3.9 & 9 & 4 & 11\\\hline
Shift in the $V_\mathrm{te(h)}$ in sub-threshold and strong inversion, $\alpha$ (unit-less)& 2 & 5 & 2 & 5\\\hline
Empirical parameter for $F_\mathrm{sat}$, $\beta$ (unit-less)& 2 & 2 & 2 & 2\\\hline
Contact resistance  at 298 K, $R_\mathrm{elec(hole)}$ ($\Omega$m)& $10^{-4}$ & $2\times10^{-4}$ & $65\times10^{-4}$ & --- \\\hline
Contact ohmic resistance, $R_\mathrm{ohmic1(2)}$ ($\Omega$m) & ---& --- & --- & $5\times10^{-3}(3\times10^{-3})$\\\hline
Schottky barrier resistance coefficient, $R_\mathrm{01(2)}$ ($\Omega$m) & --- & ---& ---& 0.183(0)\\\hline
Gate voltage dependence for Schottky barrier resistance, $a_\mathrm{1(2)}$ (\SI{}{1/V})& --- & ---& ---& 6(---)\\\hline
Transition point between two different transport paths, $V_\mathrm{01(2)}$ (\SI{}{V}) & ---& ---& ---& 0(2)\\\hline
Adjustment the smoothness of $R_\mathrm{hole}$, $\gamma$ (\SI{}{V}) & ---& ---& ---& 0.17\\
\end{tabular}}
\end{ruledtabular}
\end{table}

\vspace{-15pt}
\subsection{Dataset 1}
\vspace{-10pt}
Dataset 1 corresponds to back-gated BP FETs with
$L = 1$ $\mu$m, $W = 7.32$ $\mu$m, $t_\mathrm{BP} = 7.3$ nm, $t_\mathrm{ox} = 15$ nm, with rotational angle $42^{\circ}$ (rotational angle $0^{\circ}$ is defined along the zigzag direction and $90^{\circ}$ is along the armchair direction.) 
Model fits to experimental transfer curves ($I_\mathrm{ds}$-$V_\mathrm{gs}$) and transconductance ($g_m = \partial I_\mathrm{ds}/\partial V_\mathrm{gs}$) for a broad range of $V_\mathrm{ds}$ values at 298 K are shown in Figs.~\ref{fig:IVdg6} (a) and (b). Figures~\ref{fig:IVdg6} (c) and (d) show the model fits to measured output curves ($I_\mathrm{ds}$-$V_\mathrm{ds}$) and output conductance ($g_d = \partial I_\mathrm{ds}/\partial V_\mathrm{ds}$) for a broad range of $V_\mathrm{gs}$ values at 298 K. The model provides an excellent fit to the measured data and has the required smoothness of current derivatives as expected of compact models. The maximum output current in this dataset is less than 100 A/m obtained at $V_\mathrm{gs} = V_\mathrm{ds}$ = -2.5 V.  
Due to its limited on-current, the effect of contact resistance is not discernible. As such, we are able to fit measured data with negligible contact resistance. Neglecting the contact resistance also allows us to extract values of carrier mobility and velocity that are within the expected range for this set of BP FETs. 

We further validate this dataset using the model at $T =$ 280 K, 240 K, 200 K, and 180 K. Results are shown in Fig.~\ref{fig:temp6}. For all temperatures considered here, the maximum measured current stays under 100 A/m, which can be explained by neglecting the contact resistances. The slight increase in on-current with a reduction in temperature is due to the increase in carrier saturation velocity. The dependence of $SS$ on temperature for this dataset is examined in Fig.~\ref{fig:diff_temp}. This figure shows that at low $V_\mathrm{ds}$, $SS$ decreases at lower temperature, while at high $V_\mathrm{ds}$, $SS$ is nearly independent of temperature. 
Hence, the assumption that $n_0$ is independent of temperature and that $n_d$ varies linearly with temperature (Eq. (\ref{eq:nd})) is justified.

\vspace{-15pt}
\subsection{Dataset 2}
\vspace{-10pt}
The second dataset corresponds to back-gated BP FETs with $L = 0.3$ $\mu$m, $W = 3.16$ $\mu$m, $t_\mathrm{BP} = 8.1$ nm, $t_\mathrm{ox} = 15$ nm, and rotational angle of $40^{\circ}$. 
Unlike dataset 1, where the effect of contact resistances is not evident due to the low on-current of the device, 
a proper consideration of contact resistances is required to interpret dataset 2. This is because the channel length of the device in dataset 2 is $3\times$ smaller than that of the device analyzed in dataset 1. 
Assuming a constant value for $R_\mathrm{hole}$ = 21 $\times 10^{-4}$ $\Omega$m, the fit of the model to measured $I_\mathrm{ds}-V_\mathrm{gs}$ data is shown in Fig.~\ref{fig:const_R}.
The figure clearly shows that a constant $R_\mathrm{hole}$ is inadequate to explain the experimental data.
A higher value of $R_\mathrm{hole}$ for $V_\mathrm{gs} > -2.1$ V and a lower value of $R_\mathrm{hole}$ for $V_\mathrm{gs} < -2.1$ V can significantly improve the fit quality. 
Therefore, a bias-dependent nonlinear model of $R_\mathrm{hole}$ is necessitated in this case as discussed in Sec.~\ref{sec:contact}.

Model fits to the experimental I-V, transconductance, and output conductance data for the dataset 2 are shown in Figs.~\ref{fig:IVdg7} (a)-(d). Model parameters extracted from this fit are listed in Table~\ref{tab:table_t}. 
The validation of the model at various temperatures $T = $ 280, 240, 200, and 180 K for dataset 2 is shown in Fig.~\ref{fig:temp7}. Because of the lack of experimental data at low temperatures, we only focus on $V_\mathrm{ds}\leq-1$ V. The model provides an excellent match with experimental data over broad bias and temperature range for this device.

\vspace{-15pt}
\subsection{Datasets 3 and 4}
\vspace{-10pt}
The last two datasets are obtained through numerical simulations using the TCAD tool Sentaurus for top-gated and back-gated BP FETs with channel length $L = 0.3$ $\mu$m, $W = 3.16$ $\mu$m, $t_\mathrm{BP} = 8.1$ nm, top-oxide thickness, $t_\mathrm{ox,top}$ = 10 nm and back-oxide thickness $t_\mathrm{ox,back} = 15$ nm, and source/drain contact length, $L_\mathrm{cont} = 1.16$ $\mu$m. In Sentaurus,
the Poisson and current continuity equations are solved self-consistently for both the contact
and channel regions under the drift-diffusion approximation. Since Sentaurus does not provide parameters for 2D materials, we use previously published results obtained from experimental and theoretical calculations summarized in table~\ref{tab:table_sentaurus}.~\cite{haratipour2016fundamental, penumatcha2015analysing, qiao2014high} The Schottky contact is defined as a boundary condition between the contacts and the semiconductor. The work function of the metal contact is chosen as $\phi_m = 4.046$ eV, and the electron affinity of BP is $\chi_\mathrm{BP} = 3.655$ eV.~\cite{haratipour2016fundamental} Besides the thermionic emission, thermally assisted and direct tunneling through the SB
are also considered using a non-local tunneling model. \cite{arutchelvan2017metal} In this simulation, we ignore the effects of traps and carriers generation and recombination. 

Model fits to the numerical I-V, transconductance, and output conductance data for the dataset 3 are shown in Figs.~\ref{fig:IVdg_topG} (a)-(d). 
Model fits corresponding to dataset 4 also shown an excellent agreement with the data, but are omitted for brevity.
Model parameters extracted from this fit are listed in Table~\ref{tab:table_dataset3}. 
Also, the validation of the model at various temperatures $T = $ 280 K, 240 K, 200 K, and 180 K for dataset 3 is shown in Fig.~\ref{fig:temp_topG}. The model accurately predicts the I-V characteristics over broad bias and temperature range for this dataset. The main difference between the top- and back-gated structures is that 
the contact resistances are linear in the top-gated structure. This is because the transport path for carriers does not depend on the applied bias (see Appendix C for details.)
\begin{table}
\caption{\label{tab:table_sentaurus} BP parameters used for TCAD simulation obtained from the
literature \cite{haratipour2016fundamental, penumatcha2015analysing, qiao2014high}}
\begin{ruledtabular}
{\scriptsize
\begin{tabular}{lc}

Band gap (eV) & 0.69 \\\hline
Electron effective mass (unit-less) & 0.15 (armchair direction) \\
                        &1.18 (zigzag direction) \\\hline
Hole effective mass (unit-less) & 0.14 (armchair direction) \\
                    & 0.89 (zigzag direction) \\\hline
Dielectric constant (unit-less) & 8.3 \\\hline
Electron affinity (eV) & 3.655

\end{tabular}}
\end{ruledtabular}
\end{table}

For all datasets, we observe that $\mu_\mathrm{hole}>\mu_\mathrm{elec}$, which
is in agreement with experimental results previously reported in various BP FETs.\cite{liu2015semiconducting, haratipour2016fundamental, qiao2014high, chen2017rising} The extracted values of carrier mobility, unfortunately, are notably smaller than those reported in prior work.\cite{li2014black, liu2014phosphorene,qiao2014high, xia2014rediscovering} Yet, the mobility values of these devices are very close to those extracted from $g_m$ measurements of similar devices fabricated at the University of Minnesota. \cite{haratipour2016ambipolar, haratipour2015black} The reason for low mobility values in these samples is attributed to a combination of interface scattering, remote phonon scattering from the gate dielectric and the substrate, defects, and charged impurity scattering.~\cite{ong2014anisotropic, haratipour2015black}However, recent experimental work has demonstrated that the carrier mobility in back-gated BP FETs with HfO$_{2}$ dielectric can be improved to 165 \SI{}{cm^2/Vs} at room-temperature by optimizing the fabrication method.\cite{haratipour2018mobility}
The extracted values of the saturation velocity of carriers are also in agreement with those reported in prior work.\cite{chen2018large, chandrasekar2016optical,zhu2016black} As a result of the large effective mass of electrons in BP, the model correctly predicts that $v_{\mathrm{x0},e}<v_{\mathrm{x0},h}$.~\cite{chen2018large}

\vspace{-15pt}
\section{Conclusion}
\vspace{-10pt}
In this work, we develop a virtual-source-based analytic model to describe ambipolar current conduction in BP transistors over a broad range of bias and temperature values. The model comprehends the nonlinearity and the bias-dependence of Schottky source/drain contacts, which is necessary to explain the I-V behavior of short-channel back-gated BP transistors. The model can also capture the low-temperature transport in BP transistors. To accomplish this, key parameters such as the carrier mobility, saturation velocity, punch-through factor, and threshold voltage are modeled as simple temperature-dependent functions. The accuracy of the model is demonstrated by applying to top- and back-gated BP transistors with channel lengths of 1000 nm and 300 nm with temperature from 300 K to 180 K. The smoothness of the current and its derivatives is guaranteed in the model, thus satisfying a key criteria for compact models.
Since the model is based on threshold-voltage-based current calculation, it does not require a self-consistent solution based on surface potential. As such, the model is computationally less expensive and suitable to simulate and optimize BP-based circuits.

\section{ACKNOWLEDGMENTS}
Elahe Yarmoghaddam and Shaloo Rakheja acknowledge the funding support of National Science Foundation (NSF) through the grant no. CCF1565656. S. Rakheja also acknowledges the funding support of NYU Wireless through the Infistrial Affiliates Program. Nazila Haratipour and Steven J. Koester acknowledge primary support from NSF through the University of Minnesota MRSEC under Award DMR-1420013.

\appendix
\renewcommand \thesubsection{\Roman{subsection}}
\vspace{-15pt}
\section{Extraction methodology of model parameters}
\vspace{-10pt}
The parameter extraction begins by identifying the Dirac point (minimum conductivity), $V_\mathrm{min0}$, at $V_\mathrm{ds}$ = 0 V. The values of $\Delta_\mathrm{1e(h)}$, $\Delta_\mathrm{2e(h)}$, and $\Delta_\mathrm{3e(h)}$ are determined by matching the Dirac point from experimentally measured transfer curves and the model at all $V_\mathrm{ds}$ values.
As shown in Fig.~\ref{fig:IVdg6}(a) in Sec. IVA, the Dirac point voltage is a strong function of $V_\mathrm{ds}$ which varies from $0.06$ V for $V_\mathrm{ds} = -0.1$ V to $-1.14$ V for $V_\mathrm{ds} = -2.5$ V. 

\begin{figure*}
 \vspace{10pt}
{\textbf{Dataset1: $L = 1$ $\mu$m, $W = 7.32$ $\mu$m, $t_\mathrm{BP} = 7.3$ nm, $t_\mathrm{ox} = 15$ nm, with rotational angle $42^{\circ}$, and $T = 298$ K}}
 \vspace{-5pt}
     \includegraphics[width=1\linewidth]{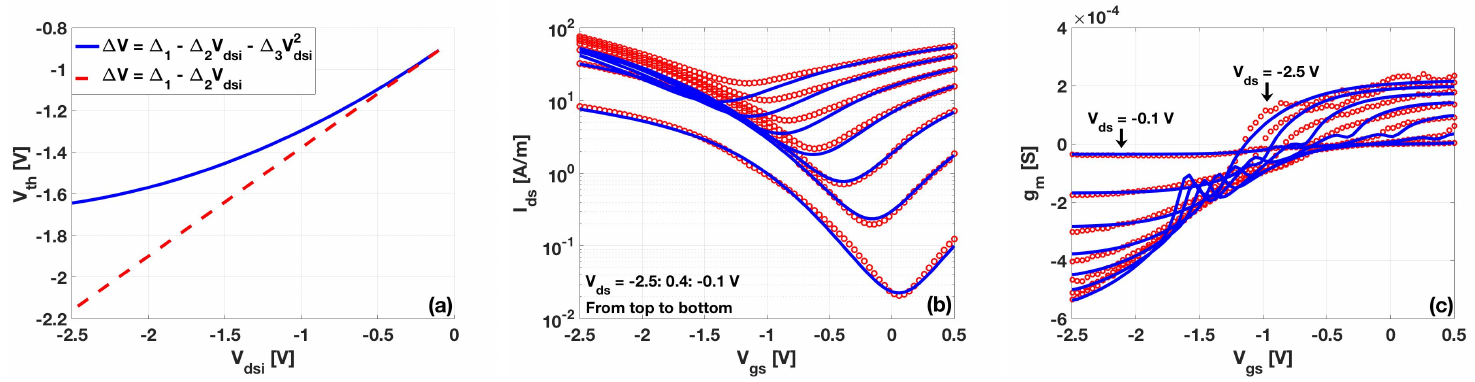}
   \caption{\label{fig:Vt} (a) hole threshold voltage ($V_\mathrm{th}$) for first dataset obtained from the model, at room temperature, by using Eq.~(\ref{eq:delta_final}) for $\Delta_\mathrm{3h}\neq0$ (solid line), and $\Delta_\mathrm{3h} = 0$ (dashed line), (b) transfer characteristics ($I_\mathrm{ds}$-$V_\mathrm{gs}$) for dataset 1, at room temperature, by setting $\Delta_\mathrm{3h} = 0$ in Eq.~(\ref{eq:delta_final}), (c) transfer conductance ($g_m = \partial I_\mathrm{ds}/\partial V_\mathrm{gs}$) for dataset 1 with $\alpha_{e/h}' = 0.5\alpha_{e/h}$ at room temperature. Solid lines are model fits, while symbols correspond to experimental data.}
\vspace{-5pt} 
 \end{figure*}

The on-off current ratio is about $400$ at $V_\mathrm{ds} = -0.1$ V. However, the on-off current ratio is as low as 4 at $V_\mathrm{ds} = -2.5$ V. This implies that the device is nearly always on at high $V_\mathrm{ds}$. This behavior can be explained by observing that even when the current due to hole conduction drops, the electron branch current increases, preventing the device from turning off at high $V_\mathrm{ds}$. By using an appropriate model of the electron and hole threshold voltages as in Eq.~\ref{eq:delta_final}, we can capture the on-off device behavior in both equilibrium and off-equilibrium transport conditions. With $\Delta_\mathrm{3h} \neq 0$ (positive value), we can model the decrease (increase) in $V_\mathrm{te(h)}$
at large negative $V_\mathrm{ds}$ values.
Figure.~\ref{fig:Vt}(a)
plots the hole threshold voltage ($V_\mathrm{th}$) predicted by the model for the first dataset as a function of $V_\mathrm{dsi}$ for $\Delta_\mathrm{3h}\neq0$ (solid line), and $\Delta_\mathrm{3h} = 0$ (dashed line). With $\Delta_\mathrm{3h} = 0$, we see that $V_\mathrm{th}$ monotonically decreases with $V_\mathrm{dsi}$, which is not the desired behavior. The transfer characteristics of the device using $\Delta_\mathrm{3h} = 0$ and other parameters as listed in Table~\ref{tab:table_t} are plotted in Fig.~\ref{fig:Vt}(b). The figure shows a poor match between the model and experimental data at highly negative $V_\mathrm{ds}$.

The transfer curve experimental data is used to obtain the values of the
non-ideality factor ($n_0$) at $V_\mathrm{ds}$ = 0 V. Similarly, the punch-through factor ($n_d$) can be obtained from experimental data by measuring the sub-threshold slope (SS) at various $V_\mathrm{ds}$ values. 
Moreover, the effective mobility values of these devices are chosen very close to those extracted from $g_m$ measurements of similar devices fabricated at the University of Minnesota. \cite{haratipour2016ambipolar, haratipour2015black}
The extracted values of the saturation velocity of carriers are also in agreement with those reported in prior work.\cite{chen2018large, chandrasekar2016optical,zhu2016black}

In this model, $\beta_{e/h}$ and $\alpha_{e/h}$ are empirical fitting parameters, and according to previous published works for VS model,~\cite{li2018analytic, khakifirooz2009simple, rakheja2014ambipolar} their values are tuned within a range of $1.5$ to $2$ and $2$ to $6$, respectively. 
Prior VS models have shown that the parameter $\alpha_{e/h}'$ is of the same order of magnitude as $\alpha_{e/h}$.  
Figure~\ref{fig:Vt} (c) shows the transconductance of dataset 1 by using $\alpha_{e/h}' = 0.5\alpha_{e/h}$, which is the typical value of $\alpha_{e/h}'$ in prior VS models. All other fitting parameters are the same as those listed in Table~\ref{tab:table_t}. 
This figure shows that for $\alpha_{e/h}' = 0.5\alpha_{e/h}$, the transconductance displays several kinks, which can be smoothed by using a slightly larger value of $\alpha_{e/h}'$ as listed in Table~\ref{tab:table_t}.

The value of $R_\mathrm{elec}$ is obtained using TLM measurements reported in Refs.~\onlinecite{haratipour2017high, haratipour2015black}.
The process to find optimal parameters in $R_\mathrm{hole}$, which is non-linear and bias-dependent, is as follows. First, we choose a large negative value of $V_\mathrm{gs}$ such that the term $\exp(aV_\mathrm{gsi}) \rightarrow 0$ (see Eq. (\ref{multline:R})). This allows us to extract appropriate values of $R_\mathrm{ohmic,1(2)}$ and $\gamma$ using the measured output characteristics. 
On the other hand, at very low $|V_\mathrm{gs}|$, $\exp{\left(aV_\mathrm{gsi}\right)}$ in Eq. (\ref{multline:R}) approaches unity. Using experimental I-V data in this regime, we can identify the value of $R_\mathrm{01(2)}$. The parameters $a_\mathrm{1(2)}$ and $V_{01(2)}$ are empirical in nature and extracted by minimizing the least-square error between the model and experimental I-V data. Finally, we use the I-V measurements at different temperatures to identify the temperature dependence of key model parameters, namely mobility, saturation velocity, punch-through factor, and threshold voltage. The extracted temperature coefficients lie in the expected range based on previous experimental and theoretical predictions.

\vspace{-15pt} 
\section{Carrier concentration dependent mobility model}\label{sec:mob}
\vspace{-10pt} 

In recent experimental work, the effect of carrier concentration on both hole and electron mobility in back-gated BP FETs was explored over a broad range of temperatures.\cite{haratipour2018mobility} As a result of electrostatic screening in the sample, the carrier mobility was found to increase with an increase in carrier concentration for all temperatures ranging from 77 K to 295 K. To handle the variation in carrier mobility with carrier concentration, Eq.(\ref{eq:mu}) is modified as
\begin{equation}
 \mu_{e/h} = \mu_{0,e/h}\left(1 + \dfrac{Q_\mathrm{x0,e/h}}{Q_\mathrm{0,e/h}}\right)^{B_{e/h}}\left(\frac{298}{T}\right)^{\xi_{\mu,{e/h}}},
 \label{eq:mu_mod}
 \end{equation}
where $\mu_{0,e/h}$, $B_{e/h}$, and $Q_\mathrm{0,e/h}$ are fitting parameters depending on the oxide dielectric constant, BP crystal orientation, and impurity density,\cite{ma2014charge} $Q_\mathrm{x0,e/h}$ is also given in Eq.~\ref{eq:Q_compact}.

\begin{figure}[h!]
\vspace{-5pt}
{Dataset 2: $L = 0.3$ $\mu$m, $W = 3.16$ $\mu$m, $t_\mathrm{BP} = 8.1$ nm, $t_\mathrm{ox} = 15$ nm, with rotational angle of $40^{\circ}$}
\vspace{-5pt}
\includegraphics [width=8.6cm, height=6.5cm]{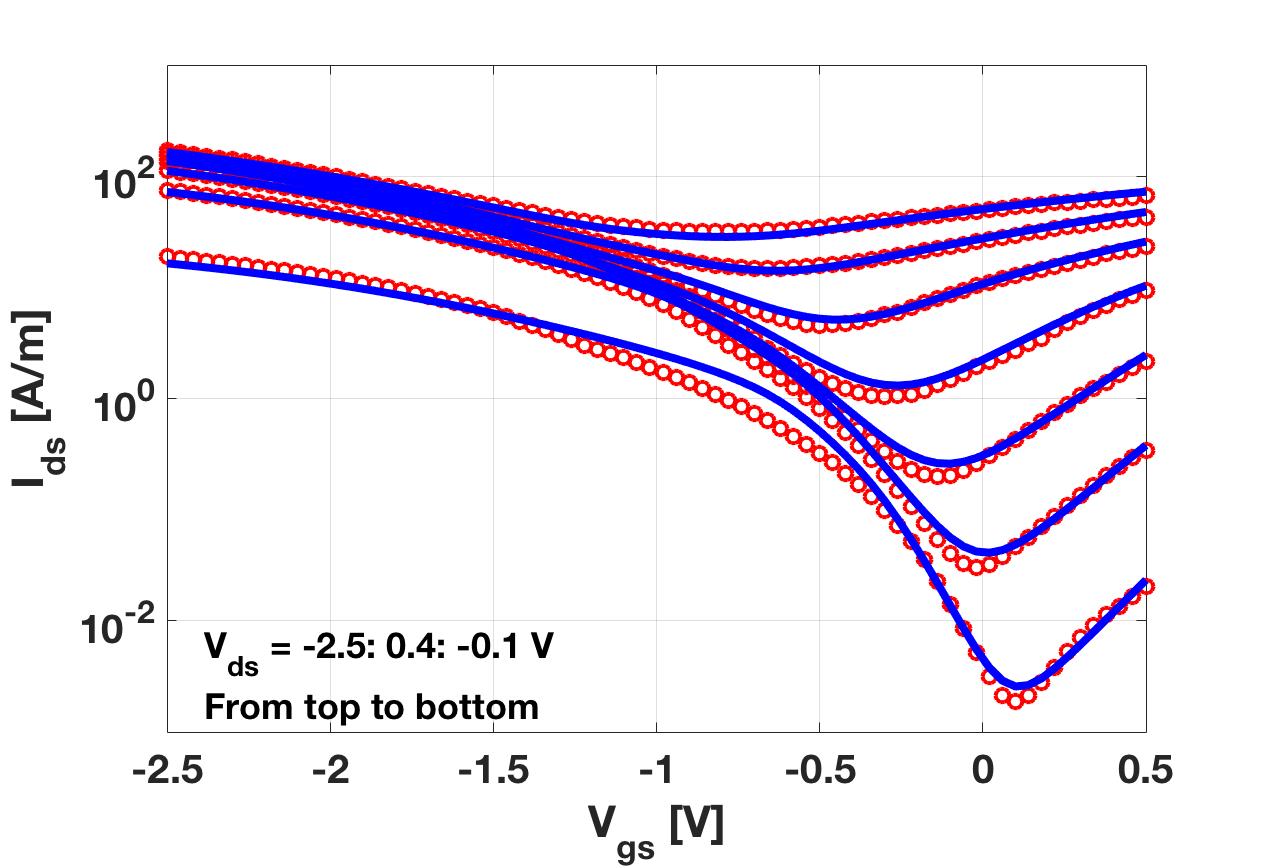}
\vspace{-5pt}
\caption{\label{fig:IVg7_mu_Q} Transfer characteristics ($I_\mathrm{ds}$-$V_\mathrm{gs}$) for dataset 2, at room temperature, by using Eq.\ref{eq:mu_mod} for hole mobility. $\mu_{0,e/h} = 60$ cm$^2$/Vs, $B_{e/h} = 1.5$, $Q_\mathrm{0,e/h} = 0.03$ C/m$^2$, $R_\mathrm{ohmic1(2)} = 0.0016(0.0017)$ $\Omega$-m, $R_\mathrm{01(2)} = 0.1155(0.0146)$ $\Omega$-m, $a_\mathrm{1(2)} = 2.0972(1.4910)$ 1/V, $\Delta_\mathrm{1h(2h)} = 0.69(0.81)$ V(1/V), and $\alpha' = 4.7$, all other fitting parameters are equal to the values mentioned in Table~\ref{tab:table_t}. Solid lines are model fits, while symbols correspond to numerical data.}
\vspace{-5pt}
\end{figure}

The updated model introduces an additional three fitting parameters, which can be determined from measurement data such as in Ref.~\citenum{haratipour2018mobility}. Model parameters corresponding to dataset 2 at 298 K are extracted using Eq.\ref{eq:mu_mod} for the hole mobility. The fitting results are shown in Fig.~\ref{fig:IVg7_mu_Q}. Here, only the parameters corresponding to the contact resistances ($R_\mathrm{ohmic1(2)}$, $R_\mathrm{01(2)}$, and $a_\mathrm{1(2)}$), shift in threshold voltage ($\Delta_\mathrm{1h(2h)}$), and empirical parameter $\alpha'$ are tweaked, while the remainder model parameters are the same as those reported in Table~\ref{tab:table_t}. With a positive value of $B_{e/h}$, the carrier mobility increases at higher $V_\mathrm{gs}$ (higher carrier concentration), necessitating a slightly larger value of the hole contact resistance to match the experimental $I_\mathrm{ds}$ data in on-state.
 
\vspace{-15pt} 
\section{Electrostatic potential and contact resistances}\label{sec:sent}
\begin{figure*}
    \centering\includegraphics[width=16cm]{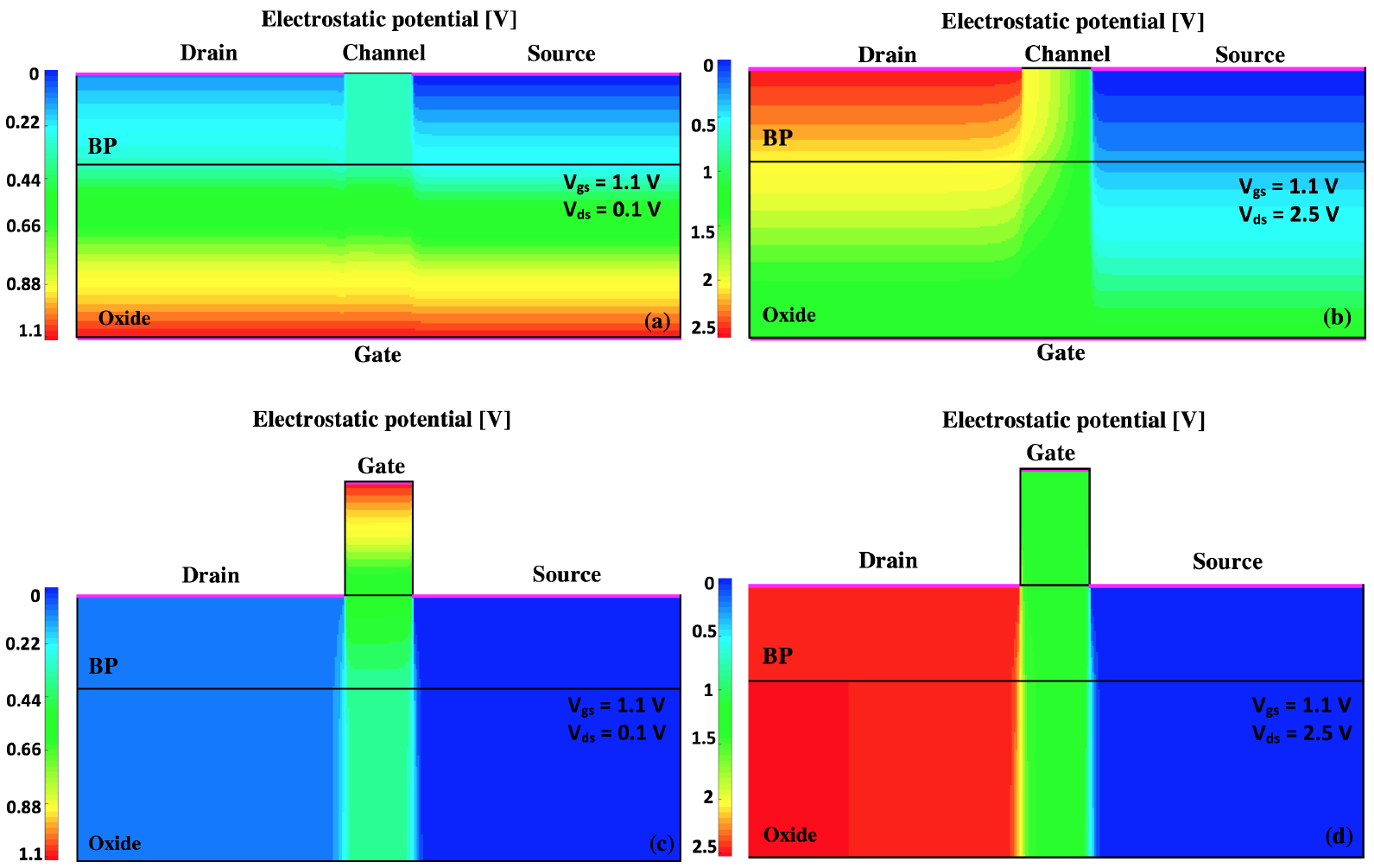}
  \caption{\label{fig:pot_sen} Surface plot of the electrostatic potential for the back-gated simulated device with $L_\mathrm{ch} = 0.3$ $\mu$m, $W = 3.16$ $\mu$m, $t_\mathrm{BP} = 8.1$ nm, $t_\mathrm{ox} = 15$ nm, and $L_\mathrm{cont} = 1.16$ $\mu$m (dataset 4) for (a) $V_\mathrm{gs} = 1.1$ V and $V_\mathrm{ds}$ = 0.1 V and (b) $V_\mathrm{gs} = 1.1$ V and $V_\mathrm{ds}$ = 2.5 V, and top-gated structure (dataset 3) with $L = 0.3$ $\mu$m, $W = 3.16$ $\mu$m, $t_\mathrm{BP} = 8.1$ nm, $t_\mathrm{ox, back} = 15$ nm, $t_\mathrm{ox,top} = 10$ nm, and $L_\mathrm{cont} = 1.16$ $\mu$m for (c) $V_\mathrm{gs} = 1.1$ V and $V_\mathrm{ds}$ = 0.1 V and (d) $V_\mathrm{gs} = 1.1$ V and $V_\mathrm{ds}$ = 2.5 V.}
\end{figure*}

\vspace{-10pt} 
To understand the nonlinearity of contact resistances, we consider the data obtained from numerical simulations using Synopsys Sentaurus (datasets 3 and 4 in the main text).
In Fig.~\ref{fig:pot_sen}, electrostatic potential distribution for $V_\mathrm{gs} = 1.1$, $V_\mathrm{ds}$ = 0.1 V and 2.5 V is shown. Results in Fig.~\ref{fig:pot_sen}(a) and (b) show that the carriers paths between the source/drain contacts depend on the applied bias voltages for the back-gated device. This is in agreement with the discussion in 
Sec.~\ref{sec:contact}.
For $V_\mathrm{ds}<V_\mathrm{gs}$, only paths 1 and 2 labeled in Fig.~\ref{fig:path} are transparent for current conduction. 
However, additional conduction paths for both carrier types contribute to current when $V_\mathrm{ds}>V_\mathrm{gs}$. 
On the other hand, for top-gated structure, as shown in Figs.~\ref{fig:pot_sen}(c) and (d), carrier transport is  independent of the bias voltages.

\begin{figure*}
{Numerical simulation: $L_\mathrm{ch} = 0.3$ $\mu$m, $W = 3.16$ $\mu$m, $t_\mathrm{BP} = 8.1$ nm, $t_\mathrm{ox} = 15$ nm, and contact length $L_\mathrm{cont} = 1.16$ $\mu$m.}
\vspace{-10pt}  
    \centering\includegraphics[width=16cm]{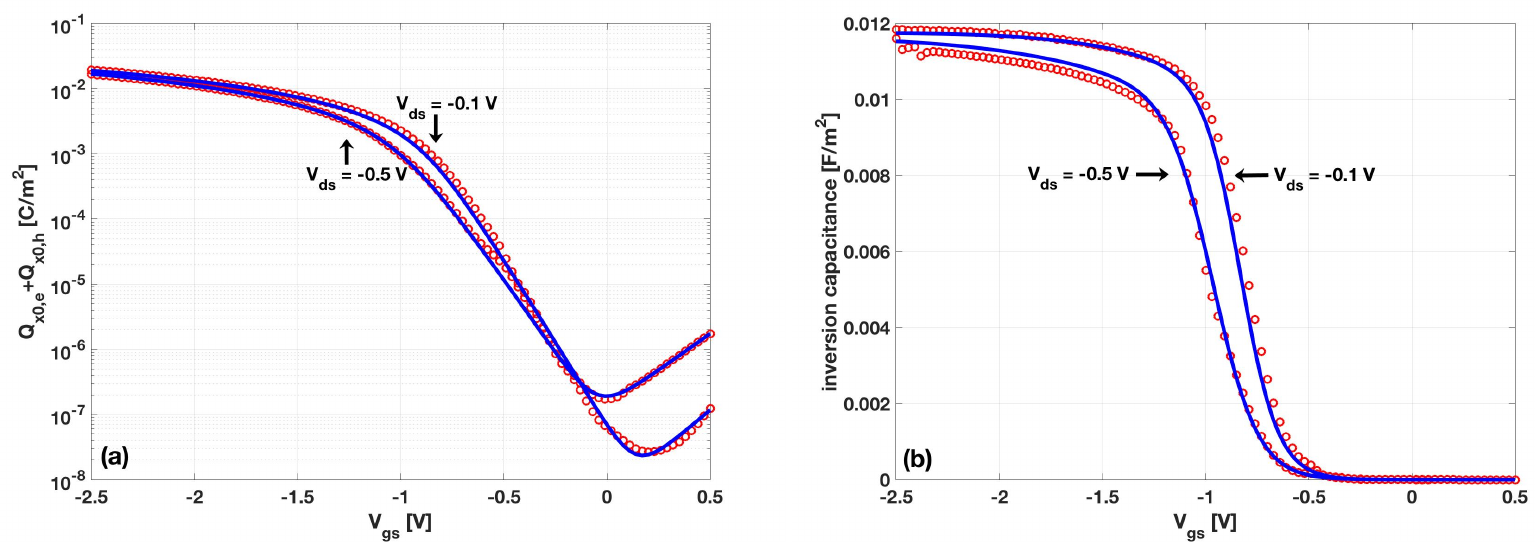}
  \caption{\label{fig:Qsen} Model fit to the numerical data obtained from TCAD simulation at room temperature for $V_\mathrm{ds} = -0.1, -0.5$ V: (a) total charge ($Q_{\mathrm{x0},e}+Q_{\mathrm{x0},h}$) as a function of $V_\mathrm{gs}$, (b) inversion capacitance (-$\partial (Q_{\mathrm{x0},e}+Q_{\mathrm{x0},h})/\partial V_\mathrm{gs}$) as a function of $V_\mathrm{gs}$. Solid lines are model fits, while symbols correspond to numerical data.}
  \vspace{-10pt}
\end{figure*}

The analytic charge model (Eq.~\ref{eq:Q_compact} in Sec. IIA) is validated by comparing the results against the charges obtained numerically for dataset 4. Results are shown in Fig.~\ref{fig:Qsen}. 
The model also faithfully reproduces the behavior of inversion capacitance  (-$\partial (Q_{\mathrm{x0},e}+Q_{\mathrm{x0},h})/\partial V_\mathrm{gs}$) versus $V_\mathrm{gs}$, thus validating our charge modeling approach.

\vspace{-15pt}
\section{Verilog-A and circuit simulation}
\begin{figure*}
\vspace{-5pt}
     \includegraphics[width=1\linewidth]{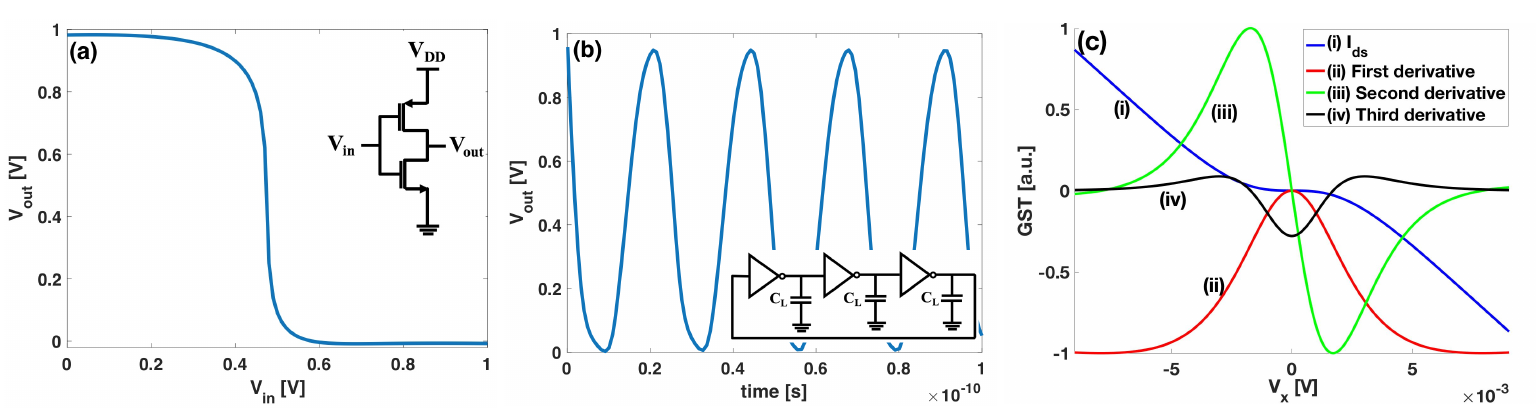}
\caption{\label{fig:verilog} (a) DC simulation of inverter: output voltage as a function of input voltage for $V_\mathrm{DD} = 1$ V. Inset shows schematic of an
inverter, (b) transient simulation for 3-stage ring oscillator where $C_L = 3$ fF. Inset indicates the schematic of the 3-stage ring oscillator, (c) results of the Gummel symmetry test (GST) for the (i) current, (ii) first, (iii) second, and (iv) third order derivatives of current with respect to $V_x$ for $V_\mathrm{gs} = 1$ V with arbitrary units. Transistor parameters are same as dataset 2.}
\vspace{-5pt}
\end{figure*}

\vspace{-10pt}
The model is implemented in Verilog-A to perform circuit simulations in SPICE.
In Fig.~\ref{fig:verilog}, the model is used to simulate the dc behavior of a BP-based inverter and the transient behavior of a BP-based 3-stage ring oscillator circuit. Transistor parameters corresponding to dataset 2 are used for these simulations. The results show the mathematical robustness of the model, ease of computation, and the capability to handle large-scale circuit simulations.

The model developed in this paper also satisfies the Gummel symmetry test (GST) required for physically accurate and well-behaved compact models. According to the GST, the model must have a symmetric formulation around $V_\mathrm{ds} = 0$ V.
Additionally, the higher order derivatives of current must be continuous.~\cite{radhakrishna2016modeling} To demonstrate that the model passes the GST, the source and drain are biased differentially ($V_x$ and $-V_x$), while the gate terminal has a fixed voltage.
Results of the GST reported in Fig.~\ref{fig:verilog} (c) verify that the model and its higher order derivatives are symmetric around $V_x = 0$ V.

\FloatBarrier
\bibliography{references}

\begin{thebibliography}{62}
\expandafter\ifx\csname natexlab\endcsname\relax\def\natexlab#1{#1}\fi
\expandafter\ifx\csname bibnamefont\endcsname\relax
  \def\bibnamefont#1{#1}\fi
\expandafter\ifx\csname bibfnamefont\endcsname\relax
  \def\bibfnamefont#1{#1}\fi
\expandafter\ifx\csname citenamefont\endcsname\relax
  \def\citenamefont#1{#1}\fi
\expandafter\ifx\csname url\endcsname\relax
  \def\url#1{\texttt{#1}}\fi
\expandafter\ifx\csname urlprefix\endcsname\relax\def\urlprefix{URL }\fi
\providecommand{\bibinfo}[2]{#2}
\providecommand{\eprint}[2][]{\url{#2}}

\bibitem[{\citenamefont{Yarmoghaddam and
  Rakheja}(2017)}]{yarmoghaddam2017dispersion}
\bibinfo{author}{\bibfnamefont{E.}~\bibnamefont{Yarmoghaddam}}
  \bibnamefont{and} \bibinfo{author}{\bibfnamefont{S.}~\bibnamefont{Rakheja}},
  \bibinfo{journal}{Journal of Applied Physics} \textbf{\bibinfo{volume}{122}},
  \bibinfo{pages}{083101} (\bibinfo{year}{2017}).

\bibitem[{\citenamefont{Rakheja
  et~al.}(2013{\natexlab{a}})\citenamefont{Rakheja, Kumar, and
  Naeemi}}]{rakheja2013evaluation}
\bibinfo{author}{\bibfnamefont{S.}~\bibnamefont{Rakheja}},
  \bibinfo{author}{\bibfnamefont{V.}~\bibnamefont{Kumar}}, \bibnamefont{and}
  \bibinfo{author}{\bibfnamefont{A.}~\bibnamefont{Naeemi}},
  \bibinfo{journal}{Proceedings of the IEEE} \textbf{\bibinfo{volume}{101}},
  \bibinfo{pages}{1740} (\bibinfo{year}{2013}{\natexlab{a}}).

\bibitem[{\citenamefont{Novoselov et~al.}(2005)\citenamefont{Novoselov, Geim,
  Morozov, Jiang, Katsnelson, Grigorieva, Dubonos, Firsov, and
  AA}}]{novoselov2005two}
\bibinfo{author}{\bibfnamefont{K.~S.} \bibnamefont{Novoselov}},
  \bibinfo{author}{\bibfnamefont{A.~K.} \bibnamefont{Geim}},
  \bibinfo{author}{\bibfnamefont{S.}~\bibnamefont{Morozov}},
  \bibinfo{author}{\bibfnamefont{D.}~\bibnamefont{Jiang}},
  \bibinfo{author}{\bibfnamefont{M.}~\bibnamefont{Katsnelson}},
  \bibinfo{author}{\bibfnamefont{I.}~\bibnamefont{Grigorieva}},
  \bibinfo{author}{\bibfnamefont{S.}~\bibnamefont{Dubonos}},
  \bibinfo{author}{\bibnamefont{Firsov}}, \bibnamefont{and}
  \bibinfo{author}{\bibnamefont{AA}}, \bibinfo{journal}{nature}
  \textbf{\bibinfo{volume}{438}}, \bibinfo{pages}{197} (\bibinfo{year}{2005}).

\bibitem[{\citenamefont{Wang et~al.}(2012)\citenamefont{Wang, Kalantar-Zadeh,
  Kis, Coleman, and Strano}}]{wang2012electronics}
\bibinfo{author}{\bibfnamefont{Q.~H.} \bibnamefont{Wang}},
  \bibinfo{author}{\bibfnamefont{K.}~\bibnamefont{Kalantar-Zadeh}},
  \bibinfo{author}{\bibfnamefont{A.}~\bibnamefont{Kis}},
  \bibinfo{author}{\bibfnamefont{J.~N.} \bibnamefont{Coleman}},
  \bibnamefont{and} \bibinfo{author}{\bibfnamefont{M.~S.}
  \bibnamefont{Strano}}, \bibinfo{journal}{Nature nanotechnology}
  \textbf{\bibinfo{volume}{7}}, \bibinfo{pages}{699} (\bibinfo{year}{2012}).

\bibitem[{\citenamefont{Fivaz and Mooser}(1967)}]{fivaz1967mobility}
\bibinfo{author}{\bibfnamefont{R.}~\bibnamefont{Fivaz}} \bibnamefont{and}
  \bibinfo{author}{\bibfnamefont{E.}~\bibnamefont{Mooser}},
  \bibinfo{journal}{Physical Review} \textbf{\bibinfo{volume}{163}},
  \bibinfo{pages}{743} (\bibinfo{year}{1967}).

\bibitem[{\citenamefont{Houssa et~al.}(2011)\citenamefont{Houssa, Scalise,
  Sankaran, Pourtois, Afanas’~Ev, and Stesmans}}]{houssa2011electronic}
\bibinfo{author}{\bibfnamefont{M.}~\bibnamefont{Houssa}},
  \bibinfo{author}{\bibfnamefont{E.}~\bibnamefont{Scalise}},
  \bibinfo{author}{\bibfnamefont{K.}~\bibnamefont{Sankaran}},
  \bibinfo{author}{\bibfnamefont{G.}~\bibnamefont{Pourtois}},
  \bibinfo{author}{\bibfnamefont{V.}~\bibnamefont{Afanas’~Ev}},
  \bibnamefont{and} \bibinfo{author}{\bibfnamefont{A.}~\bibnamefont{Stesmans}},
  \bibinfo{journal}{Applied Physics Letters} \textbf{\bibinfo{volume}{98}},
  \bibinfo{pages}{223107} (\bibinfo{year}{2011}).

\bibitem[{\citenamefont{Vogt et~al.}(2012)\citenamefont{Vogt, De~Padova,
  Quaresima, Avila, Frantzeskakis, Asensio, Resta, Ealet, and
  Le~Lay}}]{vogt2012silicene}
\bibinfo{author}{\bibfnamefont{P.}~\bibnamefont{Vogt}},
  \bibinfo{author}{\bibfnamefont{P.}~\bibnamefont{De~Padova}},
  \bibinfo{author}{\bibfnamefont{C.}~\bibnamefont{Quaresima}},
  \bibinfo{author}{\bibfnamefont{J.}~\bibnamefont{Avila}},
  \bibinfo{author}{\bibfnamefont{E.}~\bibnamefont{Frantzeskakis}},
  \bibinfo{author}{\bibfnamefont{M.~C.} \bibnamefont{Asensio}},
  \bibinfo{author}{\bibfnamefont{A.}~\bibnamefont{Resta}},
  \bibinfo{author}{\bibfnamefont{B.}~\bibnamefont{Ealet}}, \bibnamefont{and}
  \bibinfo{author}{\bibfnamefont{G.}~\bibnamefont{Le~Lay}},
  \bibinfo{journal}{Physical review letters} \textbf{\bibinfo{volume}{108}},
  \bibinfo{pages}{155501} (\bibinfo{year}{2012}).

\bibitem[{\citenamefont{Bianco et~al.}(2013)\citenamefont{Bianco, Butler,
  Jiang, Restrepo, Windl, and Goldberger}}]{bianco2013stability}
\bibinfo{author}{\bibfnamefont{E.}~\bibnamefont{Bianco}},
  \bibinfo{author}{\bibfnamefont{S.}~\bibnamefont{Butler}},
  \bibinfo{author}{\bibfnamefont{S.}~\bibnamefont{Jiang}},
  \bibinfo{author}{\bibfnamefont{O.~D.} \bibnamefont{Restrepo}},
  \bibinfo{author}{\bibfnamefont{W.}~\bibnamefont{Windl}}, \bibnamefont{and}
  \bibinfo{author}{\bibfnamefont{J.~E.} \bibnamefont{Goldberger}},
  \bibinfo{journal}{Acs Nano} \textbf{\bibinfo{volume}{7}},
  \bibinfo{pages}{4414} (\bibinfo{year}{2013}).

\bibitem[{\citenamefont{Xia et~al.}(2014{\natexlab{a}})\citenamefont{Xia, Wang,
  Xiao, Dubey, and Ramasubramaniam}}]{xia2014two}
\bibinfo{author}{\bibfnamefont{F.}~\bibnamefont{Xia}},
  \bibinfo{author}{\bibfnamefont{H.}~\bibnamefont{Wang}},
  \bibinfo{author}{\bibfnamefont{D.}~\bibnamefont{Xiao}},
  \bibinfo{author}{\bibfnamefont{M.}~\bibnamefont{Dubey}}, \bibnamefont{and}
  \bibinfo{author}{\bibfnamefont{A.}~\bibnamefont{Ramasubramaniam}},
  \bibinfo{journal}{Nature Photonics} \textbf{\bibinfo{volume}{8}},
  \bibinfo{pages}{899} (\bibinfo{year}{2014}{\natexlab{a}}).

\bibitem[{\citenamefont{Das et~al.}(2014)\citenamefont{Das, Demarteau, and
  Roelofs}}]{das2014ambipolar}
\bibinfo{author}{\bibfnamefont{S.}~\bibnamefont{Das}},
  \bibinfo{author}{\bibfnamefont{M.}~\bibnamefont{Demarteau}},
  \bibnamefont{and} \bibinfo{author}{\bibfnamefont{A.}~\bibnamefont{Roelofs}},
  \bibinfo{journal}{ACS nano} \textbf{\bibinfo{volume}{8}},
  \bibinfo{pages}{11730} (\bibinfo{year}{2014}).

\bibitem[{\citenamefont{Yi et~al.}(2017)\citenamefont{Yi, Yu, Zhou, Wang, and
  Chu}}]{yi2017two}
\bibinfo{author}{\bibfnamefont{Y.}~\bibnamefont{Yi}},
  \bibinfo{author}{\bibfnamefont{X.-F.} \bibnamefont{Yu}},
  \bibinfo{author}{\bibfnamefont{W.}~\bibnamefont{Zhou}},
  \bibinfo{author}{\bibfnamefont{J.}~\bibnamefont{Wang}}, \bibnamefont{and}
  \bibinfo{author}{\bibfnamefont{P.~K.} \bibnamefont{Chu}},
  \bibinfo{journal}{Materials Science and Engineering: R: Reports}
  \textbf{\bibinfo{volume}{120}}, \bibinfo{pages}{1} (\bibinfo{year}{2017}).

\bibitem[{\citenamefont{Qiao et~al.}(2014)\citenamefont{Qiao, Kong, Hu, Yang,
  and Ji}}]{qiao2014high}
\bibinfo{author}{\bibfnamefont{J.}~\bibnamefont{Qiao}},
  \bibinfo{author}{\bibfnamefont{X.}~\bibnamefont{Kong}},
  \bibinfo{author}{\bibfnamefont{Z.-X.} \bibnamefont{Hu}},
  \bibinfo{author}{\bibfnamefont{F.}~\bibnamefont{Yang}}, \bibnamefont{and}
  \bibinfo{author}{\bibfnamefont{W.}~\bibnamefont{Ji}},
  \bibinfo{journal}{Nature communications} \textbf{\bibinfo{volume}{5}},
  \bibinfo{pages}{4475} (\bibinfo{year}{2014}).

\bibitem[{\citenamefont{Geim and Novoselov}(2007)}]{geim2007rise}
\bibinfo{author}{\bibfnamefont{A.~K.} \bibnamefont{Geim}} \bibnamefont{and}
  \bibinfo{author}{\bibfnamefont{K.~S.} \bibnamefont{Novoselov}},
  \bibinfo{journal}{Nature materials} \textbf{\bibinfo{volume}{6}},
  \bibinfo{pages}{183} (\bibinfo{year}{2007}).

\bibitem[{\citenamefont{Hwang and Sarma}(2008)}]{hwang2008acoustic}
\bibinfo{author}{\bibfnamefont{E.}~\bibnamefont{Hwang}} \bibnamefont{and}
  \bibinfo{author}{\bibfnamefont{S.~D.} \bibnamefont{Sarma}},
  \bibinfo{journal}{Physical Review B} \textbf{\bibinfo{volume}{77}},
  \bibinfo{pages}{115449} (\bibinfo{year}{2008}).

\bibitem[{\citenamefont{Han et~al.}(2014)\citenamefont{Han, Stewart, Shevlin,
  Catlow, and Guo}}]{han2014strain}
\bibinfo{author}{\bibfnamefont{X.}~\bibnamefont{Han}},
  \bibinfo{author}{\bibfnamefont{H.~M.} \bibnamefont{Stewart}},
  \bibinfo{author}{\bibfnamefont{S.~A.} \bibnamefont{Shevlin}},
  \bibinfo{author}{\bibfnamefont{C.~R.~A.} \bibnamefont{Catlow}},
  \bibnamefont{and} \bibinfo{author}{\bibfnamefont{Z.~X.} \bibnamefont{Guo}},
  \bibinfo{journal}{Nano letters} \textbf{\bibinfo{volume}{14}},
  \bibinfo{pages}{4607} (\bibinfo{year}{2014}).

\bibitem[{\citenamefont{Du et~al.}(2014)\citenamefont{Du, Liu, Deng, and
  Ye}}]{du2014device}
\bibinfo{author}{\bibfnamefont{Y.}~\bibnamefont{Du}},
  \bibinfo{author}{\bibfnamefont{H.}~\bibnamefont{Liu}},
  \bibinfo{author}{\bibfnamefont{Y.}~\bibnamefont{Deng}}, \bibnamefont{and}
  \bibinfo{author}{\bibfnamefont{P.~D.} \bibnamefont{Ye}},
  \bibinfo{journal}{ACS nano} \textbf{\bibinfo{volume}{8}},
  \bibinfo{pages}{10035} (\bibinfo{year}{2014}).

\bibitem[{\citenamefont{Huang et~al.}(2017)\citenamefont{Huang, Xiao, Wang, Yu,
  Wang, Zhang, and Chu}}]{huang2017black}
\bibinfo{author}{\bibfnamefont{H.}~\bibnamefont{Huang}},
  \bibinfo{author}{\bibfnamefont{Q.}~\bibnamefont{Xiao}},
  \bibinfo{author}{\bibfnamefont{J.}~\bibnamefont{Wang}},
  \bibinfo{author}{\bibfnamefont{X.-F.} \bibnamefont{Yu}},
  \bibinfo{author}{\bibfnamefont{H.}~\bibnamefont{Wang}},
  \bibinfo{author}{\bibfnamefont{H.}~\bibnamefont{Zhang}}, \bibnamefont{and}
  \bibinfo{author}{\bibfnamefont{P.~K.} \bibnamefont{Chu}},
  \bibinfo{journal}{npj 2D Materials and Applications}
  \textbf{\bibinfo{volume}{1}}, \bibinfo{pages}{20} (\bibinfo{year}{2017}).

\bibitem[{\citenamefont{Haratipour et~al.}(2016)\citenamefont{Haratipour,
  Namgung, Oh, and Koester}}]{haratipour2016fundamental}
\bibinfo{author}{\bibfnamefont{N.}~\bibnamefont{Haratipour}},
  \bibinfo{author}{\bibfnamefont{S.}~\bibnamefont{Namgung}},
  \bibinfo{author}{\bibfnamefont{S.-H.} \bibnamefont{Oh}}, \bibnamefont{and}
  \bibinfo{author}{\bibfnamefont{S.~J.} \bibnamefont{Koester}},
  \bibinfo{journal}{ACS nano} \textbf{\bibinfo{volume}{10}},
  \bibinfo{pages}{3791} (\bibinfo{year}{2016}).

\bibitem[{\citenamefont{Yang et~al.}(2016)\citenamefont{Yang, Qiu, Si, Charnas,
  Milligan, Zemlyanov, Zhou, Du, Lin, Tsai et~al.}}]{yang2016few}
\bibinfo{author}{\bibfnamefont{L.}~\bibnamefont{Yang}},
  \bibinfo{author}{\bibfnamefont{G.}~\bibnamefont{Qiu}},
  \bibinfo{author}{\bibfnamefont{M.}~\bibnamefont{Si}},
  \bibinfo{author}{\bibfnamefont{A.}~\bibnamefont{Charnas}},
  \bibinfo{author}{\bibfnamefont{C.}~\bibnamefont{Milligan}},
  \bibinfo{author}{\bibfnamefont{D.}~\bibnamefont{Zemlyanov}},
  \bibinfo{author}{\bibfnamefont{H.}~\bibnamefont{Zhou}},
  \bibinfo{author}{\bibfnamefont{Y.}~\bibnamefont{Du}},
  \bibinfo{author}{\bibfnamefont{Y.}~\bibnamefont{Lin}},
  \bibinfo{author}{\bibfnamefont{W.}~\bibnamefont{Tsai}}, \bibnamefont{et~al.},
  in \emph{\bibinfo{booktitle}{Electron Devices Meeting (IEDM), 2016 IEEE
  International}} (\bibinfo{organization}{IEEE}, \bibinfo{year}{2016}), pp.
  \bibinfo{pages}{5--5}.

\bibitem[{\citenamefont{Bohloul et~al.}(2016)\citenamefont{Bohloul, Zhang,
  Gong, and Guo}}]{bohloul2016theoretical}
\bibinfo{author}{\bibfnamefont{S.}~\bibnamefont{Bohloul}},
  \bibinfo{author}{\bibfnamefont{L.}~\bibnamefont{Zhang}},
  \bibinfo{author}{\bibfnamefont{K.}~\bibnamefont{Gong}}, \bibnamefont{and}
  \bibinfo{author}{\bibfnamefont{H.}~\bibnamefont{Guo}},
  \bibinfo{journal}{Applied Physics Letters} \textbf{\bibinfo{volume}{108}},
  \bibinfo{pages}{033508} (\bibinfo{year}{2016}).

\bibitem[{\citenamefont{Luo et~al.}(2015)\citenamefont{Luo, Maassen, Deng, Du,
  Garrelts, Lundstrom, Peide, and Xu}}]{luo2015anisotropic}
\bibinfo{author}{\bibfnamefont{Z.}~\bibnamefont{Luo}},
  \bibinfo{author}{\bibfnamefont{J.}~\bibnamefont{Maassen}},
  \bibinfo{author}{\bibfnamefont{Y.}~\bibnamefont{Deng}},
  \bibinfo{author}{\bibfnamefont{Y.}~\bibnamefont{Du}},
  \bibinfo{author}{\bibfnamefont{R.~P.} \bibnamefont{Garrelts}},
  \bibinfo{author}{\bibfnamefont{M.~S.} \bibnamefont{Lundstrom}},
  \bibinfo{author}{\bibfnamefont{D.~Y.} \bibnamefont{Peide}}, \bibnamefont{and}
  \bibinfo{author}{\bibfnamefont{X.}~\bibnamefont{Xu}},
  \bibinfo{journal}{Nature communications} \textbf{\bibinfo{volume}{6}},
  \bibinfo{pages}{8572} (\bibinfo{year}{2015}).

\bibitem[{\citenamefont{Xia et~al.}(2014{\natexlab{b}})\citenamefont{Xia, Wang,
  and Jia}}]{xia2014rediscovering}
\bibinfo{author}{\bibfnamefont{F.}~\bibnamefont{Xia}},
  \bibinfo{author}{\bibfnamefont{H.}~\bibnamefont{Wang}}, \bibnamefont{and}
  \bibinfo{author}{\bibfnamefont{Y.}~\bibnamefont{Jia}},
  \bibinfo{journal}{Nature communications} \textbf{\bibinfo{volume}{5}},
  \bibinfo{pages}{4458} (\bibinfo{year}{2014}{\natexlab{b}}).

\bibitem[{\citenamefont{Rakheja et~al.}(2014)\citenamefont{Rakheja, Wu, Wang,
  Palacios, Avouris, and Antoniadis}}]{rakheja2014ambipolar}
\bibinfo{author}{\bibfnamefont{S.}~\bibnamefont{Rakheja}},
  \bibinfo{author}{\bibfnamefont{Y.}~\bibnamefont{Wu}},
  \bibinfo{author}{\bibfnamefont{H.}~\bibnamefont{Wang}},
  \bibinfo{author}{\bibfnamefont{T.}~\bibnamefont{Palacios}},
  \bibinfo{author}{\bibfnamefont{P.}~\bibnamefont{Avouris}}, \bibnamefont{and}
  \bibinfo{author}{\bibfnamefont{D.~A.} \bibnamefont{Antoniadis}},
  \bibinfo{journal}{IEEE Transactions on Nanotechnology}
  \textbf{\bibinfo{volume}{13}}, \bibinfo{pages}{1005} (\bibinfo{year}{2014}).

\bibitem[{\citenamefont{Penumatcha et~al.}(2015)\citenamefont{Penumatcha,
  Salazar, and Appenzeller}}]{penumatcha2015analysing}
\bibinfo{author}{\bibfnamefont{A.~V.} \bibnamefont{Penumatcha}},
  \bibinfo{author}{\bibfnamefont{R.~B.} \bibnamefont{Salazar}},
  \bibnamefont{and}
  \bibinfo{author}{\bibfnamefont{J.}~\bibnamefont{Appenzeller}},
  \bibinfo{journal}{Nature communications} \textbf{\bibinfo{volume}{6}},
  \bibinfo{pages}{8948} (\bibinfo{year}{2015}).

\bibitem[{\citenamefont{Esqueda et~al.}(2017)\citenamefont{Esqueda, Tian, Yan,
  and Wang}}]{esqueda2017transport}
\bibinfo{author}{\bibfnamefont{I.~S.} \bibnamefont{Esqueda}},
  \bibinfo{author}{\bibfnamefont{H.}~\bibnamefont{Tian}},
  \bibinfo{author}{\bibfnamefont{X.}~\bibnamefont{Yan}}, \bibnamefont{and}
  \bibinfo{author}{\bibfnamefont{H.}~\bibnamefont{Wang}},
  \bibinfo{journal}{IEEE Transactions on Electron Devices}
  \textbf{\bibinfo{volume}{64}}, \bibinfo{pages}{5163} (\bibinfo{year}{2017}).

\bibitem[{\citenamefont{Wang et~al.}(2016)\citenamefont{Wang, Peng, Wang, Xu,
  Ji, Lu, Li, Jin, and Liu}}]{wang2016surface}
\bibinfo{author}{\bibfnamefont{L.}~\bibnamefont{Wang}},
  \bibinfo{author}{\bibfnamefont{S.}~\bibnamefont{Peng}},
  \bibinfo{author}{\bibfnamefont{W.}~\bibnamefont{Wang}},
  \bibinfo{author}{\bibfnamefont{G.}~\bibnamefont{Xu}},
  \bibinfo{author}{\bibfnamefont{Z.}~\bibnamefont{Ji}},
  \bibinfo{author}{\bibfnamefont{N.}~\bibnamefont{Lu}},
  \bibinfo{author}{\bibfnamefont{L.}~\bibnamefont{Li}},
  \bibinfo{author}{\bibfnamefont{Z.}~\bibnamefont{Jin}}, \bibnamefont{and}
  \bibinfo{author}{\bibfnamefont{M.}~\bibnamefont{Liu}},
  \bibinfo{journal}{Journal of Applied Physics} \textbf{\bibinfo{volume}{120}},
  \bibinfo{pages}{084509} (\bibinfo{year}{2016}).

\bibitem[{\citenamefont{Wang et~al.}(2017)\citenamefont{Wang, Li, Feng, Ang,
  Gong, Thean, and Liang}}]{wang2017unified}
\bibinfo{author}{\bibfnamefont{L.}~\bibnamefont{Wang}},
  \bibinfo{author}{\bibfnamefont{Y.}~\bibnamefont{Li}},
  \bibinfo{author}{\bibfnamefont{X.}~\bibnamefont{Feng}},
  \bibinfo{author}{\bibfnamefont{K.-W.} \bibnamefont{Ang}},
  \bibinfo{author}{\bibfnamefont{X.}~\bibnamefont{Gong}},
  \bibinfo{author}{\bibfnamefont{A.}~\bibnamefont{Thean}}, \bibnamefont{and}
  \bibinfo{author}{\bibfnamefont{G.}~\bibnamefont{Liang}}, in
  \emph{\bibinfo{booktitle}{Electron Devices Meeting (IEDM), 2017 IEEE
  International}} (\bibinfo{organization}{IEEE}, \bibinfo{year}{2017}), pp.
  \bibinfo{pages}{31--4}.

\bibitem[{\citenamefont{Cao et~al.}(2018)\citenamefont{Cao, Peng, Liu, Wu, Li,
  Geng, Yang, Ji, Lu, and Liu}}]{cao2018new}
\bibinfo{author}{\bibfnamefont{J.}~\bibnamefont{Cao}},
  \bibinfo{author}{\bibfnamefont{S.}~\bibnamefont{Peng}},
  \bibinfo{author}{\bibfnamefont{W.}~\bibnamefont{Liu}},
  \bibinfo{author}{\bibfnamefont{Q.}~\bibnamefont{Wu}},
  \bibinfo{author}{\bibfnamefont{L.}~\bibnamefont{Li}},
  \bibinfo{author}{\bibfnamefont{D.}~\bibnamefont{Geng}},
  \bibinfo{author}{\bibfnamefont{G.}~\bibnamefont{Yang}},
  \bibinfo{author}{\bibfnamefont{Z.}~\bibnamefont{Ji}},
  \bibinfo{author}{\bibfnamefont{N.}~\bibnamefont{Lu}}, \bibnamefont{and}
  \bibinfo{author}{\bibfnamefont{M.}~\bibnamefont{Liu}},
  \bibinfo{journal}{Journal of Applied Physics} \textbf{\bibinfo{volume}{123}},
  \bibinfo{pages}{064501} (\bibinfo{year}{2018}).

\bibitem[{\citenamefont{Taur et~al.}(2016)\citenamefont{Taur, Wu, and
  Min}}]{taur2016short}
\bibinfo{author}{\bibfnamefont{Y.}~\bibnamefont{Taur}},
  \bibinfo{author}{\bibfnamefont{J.}~\bibnamefont{Wu}}, \bibnamefont{and}
  \bibinfo{author}{\bibfnamefont{J.}~\bibnamefont{Min}}, \bibinfo{journal}{IEEE
  Transactions on Electron Devices} \textbf{\bibinfo{volume}{63}},
  \bibinfo{pages}{2550} (\bibinfo{year}{2016}).

\bibitem[{\citenamefont{Marin et~al.}(2018)\citenamefont{Marin, Bader, and
  Jena}}]{marin2018new}
\bibinfo{author}{\bibfnamefont{E.~G.} \bibnamefont{Marin}},
  \bibinfo{author}{\bibfnamefont{S.~J.} \bibnamefont{Bader}}, \bibnamefont{and}
  \bibinfo{author}{\bibfnamefont{D.}~\bibnamefont{Jena}},
  \bibinfo{journal}{IEEE Transactions on Electron Devices}
  \textbf{\bibinfo{volume}{65}}, \bibinfo{pages}{1239} (\bibinfo{year}{2018}).

\bibitem[{\citenamefont{Landauer}(1957)}]{landauer1957spatial}
\bibinfo{author}{\bibfnamefont{R.}~\bibnamefont{Landauer}},
  \bibinfo{journal}{IBM Journal of Research and Development}
  \textbf{\bibinfo{volume}{1}}, \bibinfo{pages}{223} (\bibinfo{year}{1957}).

\bibitem[{\citenamefont{Datta}(1997)}]{datta1997electronic}
\bibinfo{author}{\bibfnamefont{S.}~\bibnamefont{Datta}},
  \emph{\bibinfo{title}{Electronic transport in mesoscopic systems}}
  (\bibinfo{publisher}{Cambridge university press}, \bibinfo{year}{1997}).

\bibitem[{\citenamefont{Lundstrom and Antoniadis}(2014)}]{lundstrom2014compact}
\bibinfo{author}{\bibfnamefont{M.~S.} \bibnamefont{Lundstrom}}
  \bibnamefont{and} \bibinfo{author}{\bibfnamefont{D.~A.}
  \bibnamefont{Antoniadis}}, \bibinfo{journal}{IEEE Transactions on Electron
  Devices} \textbf{\bibinfo{volume}{61}}, \bibinfo{pages}{225}
  (\bibinfo{year}{2014}).

\bibitem[{\citenamefont{Radhakrishna et~al.}(2014)\citenamefont{Radhakrishna,
  Imada, Palacios, and Antoniadis}}]{radhakrishna2014virtual}
\bibinfo{author}{\bibfnamefont{U.}~\bibnamefont{Radhakrishna}},
  \bibinfo{author}{\bibfnamefont{T.}~\bibnamefont{Imada}},
  \bibinfo{author}{\bibfnamefont{T.}~\bibnamefont{Palacios}}, \bibnamefont{and}
  \bibinfo{author}{\bibfnamefont{D.}~\bibnamefont{Antoniadis}},
  \bibinfo{journal}{physica status solidi (c)} \textbf{\bibinfo{volume}{11}},
  \bibinfo{pages}{848} (\bibinfo{year}{2014}).

\bibitem[{\citenamefont{Rakheja
  et~al.}(2013{\natexlab{b}})\citenamefont{Rakheja, Wang, Palacios, Meric,
  Shepard, and Antoniadis}}]{rakheja2013unified}
\bibinfo{author}{\bibfnamefont{S.}~\bibnamefont{Rakheja}},
  \bibinfo{author}{\bibfnamefont{H.}~\bibnamefont{Wang}},
  \bibinfo{author}{\bibfnamefont{T.}~\bibnamefont{Palacios}},
  \bibinfo{author}{\bibfnamefont{I.}~\bibnamefont{Meric}},
  \bibinfo{author}{\bibfnamefont{K.}~\bibnamefont{Shepard}}, \bibnamefont{and}
  \bibinfo{author}{\bibfnamefont{D.}~\bibnamefont{Antoniadis}}, in
  \emph{\bibinfo{booktitle}{Electron Devices Meeting (IEDM), 2013 IEEE
  International}} (\bibinfo{organization}{IEEE},
  \bibinfo{year}{2013}{\natexlab{b}}), pp. \bibinfo{pages}{5--5}.

\bibitem[{\citenamefont{Haratipour et~al.}(2015)\citenamefont{Haratipour,
  Robbins, and Koester}}]{haratipour2015black}
\bibinfo{author}{\bibfnamefont{N.}~\bibnamefont{Haratipour}},
  \bibinfo{author}{\bibfnamefont{M.~C.} \bibnamefont{Robbins}},
  \bibnamefont{and} \bibinfo{author}{\bibfnamefont{S.~J.}
  \bibnamefont{Koester}}, in \emph{\bibinfo{booktitle}{Device Research
  Conference (DRC), 2015 73rd Annual}} (\bibinfo{organization}{IEEE},
  \bibinfo{year}{2015}), pp. \bibinfo{pages}{243--244}.

\bibitem[{\citenamefont{Haratipour and
  Koester}(2016)}]{haratipour2016ambipolar}
\bibinfo{author}{\bibfnamefont{N.}~\bibnamefont{Haratipour}} \bibnamefont{and}
  \bibinfo{author}{\bibfnamefont{S.~J.} \bibnamefont{Koester}},
  \bibinfo{journal}{IEEE Electron Device Letters}
  \textbf{\bibinfo{volume}{37}}, \bibinfo{pages}{103} (\bibinfo{year}{2016}).

\bibitem[{\citenamefont{Chang et~al.}(2014)\citenamefont{Chang, Zhu, and
  Akinwande}}]{chang2014mobility}
\bibinfo{author}{\bibfnamefont{H.-Y.} \bibnamefont{Chang}},
  \bibinfo{author}{\bibfnamefont{W.}~\bibnamefont{Zhu}}, \bibnamefont{and}
  \bibinfo{author}{\bibfnamefont{D.}~\bibnamefont{Akinwande}},
  \bibinfo{journal}{Applied Physics Letters} \textbf{\bibinfo{volume}{104}},
  \bibinfo{pages}{113504} (\bibinfo{year}{2014}).

\bibitem[{\citenamefont{Chang et~al.}(2017)\citenamefont{Chang, Charnas, Lin,
  Peide, Wu, and Wu}}]{chang2017germanium}
\bibinfo{author}{\bibfnamefont{H.-M.} \bibnamefont{Chang}},
  \bibinfo{author}{\bibfnamefont{A.}~\bibnamefont{Charnas}},
  \bibinfo{author}{\bibfnamefont{Y.-M.} \bibnamefont{Lin}},
  \bibinfo{author}{\bibfnamefont{D.~Y.} \bibnamefont{Peide}},
  \bibinfo{author}{\bibfnamefont{C.-I.} \bibnamefont{Wu}}, \bibnamefont{and}
  \bibinfo{author}{\bibfnamefont{C.-H.} \bibnamefont{Wu}},
  \bibinfo{journal}{Scientific reports} \textbf{\bibinfo{volume}{7}},
  \bibinfo{pages}{16857} (\bibinfo{year}{2017}).

\bibitem[{\citenamefont{Du et~al.}(2016)\citenamefont{Du, Neal, Zhou, and
  Peide}}]{du2016transport}
\bibinfo{author}{\bibfnamefont{Y.}~\bibnamefont{Du}},
  \bibinfo{author}{\bibfnamefont{A.~T.} \bibnamefont{Neal}},
  \bibinfo{author}{\bibfnamefont{H.}~\bibnamefont{Zhou}}, \bibnamefont{and}
  \bibinfo{author}{\bibfnamefont{D.~Y.} \bibnamefont{Peide}},
  \bibinfo{journal}{Journal of Physics: Condensed Matter}
  \textbf{\bibinfo{volume}{28}}, \bibinfo{pages}{263002}
  (\bibinfo{year}{2016}).

\bibitem[{\citenamefont{Xu et~al.}(2016)\citenamefont{Xu, Cheng, Du, Yang, Yu,
  Luo, Yin, Li, Dong, Ye et~al.}}]{xu2016contacts}
\bibinfo{author}{\bibfnamefont{Y.}~\bibnamefont{Xu}},
  \bibinfo{author}{\bibfnamefont{C.}~\bibnamefont{Cheng}},
  \bibinfo{author}{\bibfnamefont{S.}~\bibnamefont{Du}},
  \bibinfo{author}{\bibfnamefont{J.}~\bibnamefont{Yang}},
  \bibinfo{author}{\bibfnamefont{B.}~\bibnamefont{Yu}},
  \bibinfo{author}{\bibfnamefont{J.}~\bibnamefont{Luo}},
  \bibinfo{author}{\bibfnamefont{W.}~\bibnamefont{Yin}},
  \bibinfo{author}{\bibfnamefont{E.}~\bibnamefont{Li}},
  \bibinfo{author}{\bibfnamefont{S.}~\bibnamefont{Dong}},
  \bibinfo{author}{\bibfnamefont{P.}~\bibnamefont{Ye}}, \bibnamefont{et~al.},
  \bibinfo{journal}{ACS nano} \textbf{\bibinfo{volume}{10}},
  \bibinfo{pages}{4895} (\bibinfo{year}{2016}).

\bibitem[{\citenamefont{Haratipour et~al.}(2017)\citenamefont{Haratipour,
  Namgung, Grassi, Low, Oh, and Koester}}]{haratipour2017high}
\bibinfo{author}{\bibfnamefont{N.}~\bibnamefont{Haratipour}},
  \bibinfo{author}{\bibfnamefont{S.}~\bibnamefont{Namgung}},
  \bibinfo{author}{\bibfnamefont{R.}~\bibnamefont{Grassi}},
  \bibinfo{author}{\bibfnamefont{T.}~\bibnamefont{Low}},
  \bibinfo{author}{\bibfnamefont{S.-H.} \bibnamefont{Oh}}, \bibnamefont{and}
  \bibinfo{author}{\bibfnamefont{S.~J.} \bibnamefont{Koester}},
  \bibinfo{journal}{IEEE Electron Device Letters}
  \textbf{\bibinfo{volume}{38}}, \bibinfo{pages}{685} (\bibinfo{year}{2017}).

\bibitem[{\citenamefont{Valletta et~al.}(2011)\citenamefont{Valletta, Daami,
  Benwadih, Coppard, Fortunato, Rapisarda, Torricelli, and
  Mariucci}}]{valletta2011contact}
\bibinfo{author}{\bibfnamefont{A.}~\bibnamefont{Valletta}},
  \bibinfo{author}{\bibfnamefont{A.}~\bibnamefont{Daami}},
  \bibinfo{author}{\bibfnamefont{M.}~\bibnamefont{Benwadih}},
  \bibinfo{author}{\bibfnamefont{R.}~\bibnamefont{Coppard}},
  \bibinfo{author}{\bibfnamefont{G.}~\bibnamefont{Fortunato}},
  \bibinfo{author}{\bibfnamefont{M.}~\bibnamefont{Rapisarda}},
  \bibinfo{author}{\bibfnamefont{F.}~\bibnamefont{Torricelli}},
  \bibnamefont{and} \bibinfo{author}{\bibfnamefont{L.}~\bibnamefont{Mariucci}},
  \bibinfo{journal}{Applied Physics Letters} \textbf{\bibinfo{volume}{99}},
  \bibinfo{pages}{271} (\bibinfo{year}{2011}).

\bibitem[{\citenamefont{Ong et~al.}(2014)\citenamefont{Ong, Zhang, and
  Zhang}}]{ong2014anisotropic}
\bibinfo{author}{\bibfnamefont{Z.-Y.} \bibnamefont{Ong}},
  \bibinfo{author}{\bibfnamefont{G.}~\bibnamefont{Zhang}}, \bibnamefont{and}
  \bibinfo{author}{\bibfnamefont{Y.~W.} \bibnamefont{Zhang}},
  \bibinfo{journal}{Journal of Applied Physics} \textbf{\bibinfo{volume}{116}},
  \bibinfo{pages}{214505} (\bibinfo{year}{2014}).

\bibitem[{\citenamefont{Jariwala et~al.}(2014)\citenamefont{Jariwala, Sangwan,
  Lauhon, Marks, and Hersam}}]{jariwala2014emerging}
\bibinfo{author}{\bibfnamefont{D.}~\bibnamefont{Jariwala}},
  \bibinfo{author}{\bibfnamefont{V.~K.} \bibnamefont{Sangwan}},
  \bibinfo{author}{\bibfnamefont{L.~J.} \bibnamefont{Lauhon}},
  \bibinfo{author}{\bibfnamefont{T.~J.} \bibnamefont{Marks}}, \bibnamefont{and}
  \bibinfo{author}{\bibfnamefont{M.~C.} \bibnamefont{Hersam}},
  \bibinfo{journal}{ACS nano} \textbf{\bibinfo{volume}{8}},
  \bibinfo{pages}{1102} (\bibinfo{year}{2014}).

\bibitem[{\citenamefont{Radisavljevic and
  Kis}(2013)}]{radisavljevic2013mobility}
\bibinfo{author}{\bibfnamefont{B.}~\bibnamefont{Radisavljevic}}
  \bibnamefont{and} \bibinfo{author}{\bibfnamefont{A.}~\bibnamefont{Kis}},
  \bibinfo{journal}{Nature materials} \textbf{\bibinfo{volume}{12}},
  \bibinfo{pages}{815} (\bibinfo{year}{2013}).

\bibitem[{\citenamefont{Ovchinnikov et~al.}(2014)\citenamefont{Ovchinnikov,
  Allain, Huang, Dumcenco, and Kis}}]{ovchinnikov2014electrical}
\bibinfo{author}{\bibfnamefont{D.}~\bibnamefont{Ovchinnikov}},
  \bibinfo{author}{\bibfnamefont{A.}~\bibnamefont{Allain}},
  \bibinfo{author}{\bibfnamefont{Y.-S.} \bibnamefont{Huang}},
  \bibinfo{author}{\bibfnamefont{D.}~\bibnamefont{Dumcenco}}, \bibnamefont{and}
  \bibinfo{author}{\bibfnamefont{A.}~\bibnamefont{Kis}}, \bibinfo{journal}{ACS
  nano} \textbf{\bibinfo{volume}{8}}, \bibinfo{pages}{8174}
  (\bibinfo{year}{2014}).

\bibitem[{\citenamefont{Trushkov and Perebeinos}(2017)}]{trushkov2017phonon}
\bibinfo{author}{\bibfnamefont{Y.}~\bibnamefont{Trushkov}} \bibnamefont{and}
  \bibinfo{author}{\bibfnamefont{V.}~\bibnamefont{Perebeinos}},
  \bibinfo{journal}{Physical Review B} \textbf{\bibinfo{volume}{95}},
  \bibinfo{pages}{075436} (\bibinfo{year}{2017}).

\bibitem[{\citenamefont{Li et~al.}(2014)\citenamefont{Li, Yu, Ye, Ge, Ou, Wu,
  Feng, Chen, and Zhang}}]{li2014black}
\bibinfo{author}{\bibfnamefont{L.}~\bibnamefont{Li}},
  \bibinfo{author}{\bibfnamefont{Y.}~\bibnamefont{Yu}},
  \bibinfo{author}{\bibfnamefont{G.~J.} \bibnamefont{Ye}},
  \bibinfo{author}{\bibfnamefont{Q.}~\bibnamefont{Ge}},
  \bibinfo{author}{\bibfnamefont{X.}~\bibnamefont{Ou}},
  \bibinfo{author}{\bibfnamefont{H.}~\bibnamefont{Wu}},
  \bibinfo{author}{\bibfnamefont{D.}~\bibnamefont{Feng}},
  \bibinfo{author}{\bibfnamefont{X.~H.} \bibnamefont{Chen}}, \bibnamefont{and}
  \bibinfo{author}{\bibfnamefont{Y.}~\bibnamefont{Zhang}},
  \bibinfo{journal}{Nature nanotechnology} \textbf{\bibinfo{volume}{9}},
  \bibinfo{pages}{372} (\bibinfo{year}{2014}).

\bibitem[{\citenamefont{Haratipour et~al.}(2018)\citenamefont{Haratipour, Liu,
  Wu, Namgung, Ruden, Mkhoyan, Oh, and Koester}}]{haratipour2018mobility}
\bibinfo{author}{\bibfnamefont{N.}~\bibnamefont{Haratipour}},
  \bibinfo{author}{\bibfnamefont{Y.}~\bibnamefont{Liu}},
  \bibinfo{author}{\bibfnamefont{R.~J.} \bibnamefont{Wu}},
  \bibinfo{author}{\bibfnamefont{S.}~\bibnamefont{Namgung}},
  \bibinfo{author}{\bibfnamefont{P.~P.} \bibnamefont{Ruden}},
  \bibinfo{author}{\bibfnamefont{K.~A.} \bibnamefont{Mkhoyan}},
  \bibinfo{author}{\bibfnamefont{S.-H.} \bibnamefont{Oh}}, \bibnamefont{and}
  \bibinfo{author}{\bibfnamefont{S.~J.} \bibnamefont{Koester}},
  \bibinfo{journal}{IEEE Transactions on Electron Devices} pp.
  \bibinfo{pages}{1--9} (\bibinfo{year}{2018}).

\bibitem[{\citenamefont{Chen et~al.}(2018)\citenamefont{Chen, Chen, Levi,
  Houben, Deng, Yuan, Ma, Watanabe, Taniguchi, Naveh et~al.}}]{chen2018large}
\bibinfo{author}{\bibfnamefont{X.}~\bibnamefont{Chen}},
  \bibinfo{author}{\bibfnamefont{C.}~\bibnamefont{Chen}},
  \bibinfo{author}{\bibfnamefont{A.}~\bibnamefont{Levi}},
  \bibinfo{author}{\bibfnamefont{L.}~\bibnamefont{Houben}},
  \bibinfo{author}{\bibfnamefont{B.}~\bibnamefont{Deng}},
  \bibinfo{author}{\bibfnamefont{S.}~\bibnamefont{Yuan}},
  \bibinfo{author}{\bibfnamefont{C.}~\bibnamefont{Ma}},
  \bibinfo{author}{\bibfnamefont{K.}~\bibnamefont{Watanabe}},
  \bibinfo{author}{\bibfnamefont{T.}~\bibnamefont{Taniguchi}},
  \bibinfo{author}{\bibfnamefont{D.}~\bibnamefont{Naveh}},
  \bibnamefont{et~al.}, \bibinfo{journal}{ACS nano}
  \textbf{\bibinfo{volume}{12}}, \bibinfo{pages}{5003} (\bibinfo{year}{2018}).

\bibitem[{\citenamefont{Guide}(2016)}]{guide2016version}
\bibinfo{author}{\bibfnamefont{S.~D.~U.} \bibnamefont{Guide}},
  \bibinfo{journal}{Inc., Mountain View, CA}  (\bibinfo{year}{2016}).

\bibitem[{\citenamefont{Arutchelvan et~al.}(2017)\citenamefont{Arutchelvan,
  de~la Rosa, Matagne, Sutar, Radu, Huyghebaert, De~Gendt, and
  Heyns}}]{arutchelvan2017metal}
\bibinfo{author}{\bibfnamefont{G.}~\bibnamefont{Arutchelvan}},
  \bibinfo{author}{\bibfnamefont{C.~J.~L.} \bibnamefont{de~la Rosa}},
  \bibinfo{author}{\bibfnamefont{P.}~\bibnamefont{Matagne}},
  \bibinfo{author}{\bibfnamefont{S.}~\bibnamefont{Sutar}},
  \bibinfo{author}{\bibfnamefont{I.}~\bibnamefont{Radu}},
  \bibinfo{author}{\bibfnamefont{C.}~\bibnamefont{Huyghebaert}},
  \bibinfo{author}{\bibfnamefont{S.}~\bibnamefont{De~Gendt}}, \bibnamefont{and}
  \bibinfo{author}{\bibfnamefont{M.}~\bibnamefont{Heyns}},
  \bibinfo{journal}{Nanoscale} \textbf{\bibinfo{volume}{9}},
  \bibinfo{pages}{10869} (\bibinfo{year}{2017}).

\bibitem[{\citenamefont{Liu et~al.}(2015)\citenamefont{Liu, Du, Deng, and
  Peide}}]{liu2015semiconducting}
\bibinfo{author}{\bibfnamefont{H.}~\bibnamefont{Liu}},
  \bibinfo{author}{\bibfnamefont{Y.}~\bibnamefont{Du}},
  \bibinfo{author}{\bibfnamefont{Y.}~\bibnamefont{Deng}}, \bibnamefont{and}
  \bibinfo{author}{\bibfnamefont{D.~Y.} \bibnamefont{Peide}},
  \bibinfo{journal}{Chemical Society Reviews} \textbf{\bibinfo{volume}{44}},
  \bibinfo{pages}{2732} (\bibinfo{year}{2015}).

\bibitem[{\citenamefont{Chen et~al.}(2017)\citenamefont{Chen, Li, Chen, Ong,
  and Zhao}}]{chen2017rising}
\bibinfo{author}{\bibfnamefont{P.}~\bibnamefont{Chen}},
  \bibinfo{author}{\bibfnamefont{N.}~\bibnamefont{Li}},
  \bibinfo{author}{\bibfnamefont{X.}~\bibnamefont{Chen}},
  \bibinfo{author}{\bibfnamefont{W.-J.} \bibnamefont{Ong}}, \bibnamefont{and}
  \bibinfo{author}{\bibfnamefont{X.}~\bibnamefont{Zhao}}, \bibinfo{journal}{2D
  Materials} \textbf{\bibinfo{volume}{5}}, \bibinfo{pages}{014002}
  (\bibinfo{year}{2017}).

\bibitem[{\citenamefont{Liu et~al.}(2014)\citenamefont{Liu, Neal, Zhu, Luo, Xu,
  Tom{\'a}nek, and Ye}}]{liu2014phosphorene}
\bibinfo{author}{\bibfnamefont{H.}~\bibnamefont{Liu}},
  \bibinfo{author}{\bibfnamefont{A.~T.} \bibnamefont{Neal}},
  \bibinfo{author}{\bibfnamefont{Z.}~\bibnamefont{Zhu}},
  \bibinfo{author}{\bibfnamefont{Z.}~\bibnamefont{Luo}},
  \bibinfo{author}{\bibfnamefont{X.}~\bibnamefont{Xu}},
  \bibinfo{author}{\bibfnamefont{D.}~\bibnamefont{Tom{\'a}nek}},
  \bibnamefont{and} \bibinfo{author}{\bibfnamefont{P.~D.} \bibnamefont{Ye}},
  \bibinfo{journal}{ACS nano} \textbf{\bibinfo{volume}{8}},
  \bibinfo{pages}{4033} (\bibinfo{year}{2014}).

\bibitem[{\citenamefont{Chandrasekar et~al.}(2016)\citenamefont{Chandrasekar,
  Ganapathi, Bhattacharjee, Bhat, and Nath}}]{chandrasekar2016optical}
\bibinfo{author}{\bibfnamefont{H.}~\bibnamefont{Chandrasekar}},
  \bibinfo{author}{\bibfnamefont{K.~L.} \bibnamefont{Ganapathi}},
  \bibinfo{author}{\bibfnamefont{S.}~\bibnamefont{Bhattacharjee}},
  \bibinfo{author}{\bibfnamefont{N.}~\bibnamefont{Bhat}}, \bibnamefont{and}
  \bibinfo{author}{\bibfnamefont{D.~N.} \bibnamefont{Nath}},
  \bibinfo{journal}{IEEE Transactions on Electron Devices}
  \textbf{\bibinfo{volume}{63}}, \bibinfo{pages}{767} (\bibinfo{year}{2016}).

\bibitem[{\citenamefont{Zhu et~al.}(2016)\citenamefont{Zhu, Park, Yogeesh,
  McNicholas, Bank, and Akinwande}}]{zhu2016black}
\bibinfo{author}{\bibfnamefont{W.}~\bibnamefont{Zhu}},
  \bibinfo{author}{\bibfnamefont{S.}~\bibnamefont{Park}},
  \bibinfo{author}{\bibfnamefont{M.~N.} \bibnamefont{Yogeesh}},
  \bibinfo{author}{\bibfnamefont{K.~M.} \bibnamefont{McNicholas}},
  \bibinfo{author}{\bibfnamefont{S.~R.} \bibnamefont{Bank}}, \bibnamefont{and}
  \bibinfo{author}{\bibfnamefont{D.}~\bibnamefont{Akinwande}},
  \bibinfo{journal}{Nano letters} \textbf{\bibinfo{volume}{16}},
  \bibinfo{pages}{2301} (\bibinfo{year}{2016}).

\bibitem[{\citenamefont{Li and Rakheja}(2018)}]{li2018analytic}
\bibinfo{author}{\bibfnamefont{K.}~\bibnamefont{Li}} \bibnamefont{and}
  \bibinfo{author}{\bibfnamefont{S.}~\bibnamefont{Rakheja}},
  \bibinfo{journal}{Journal of Applied Physics} \textbf{\bibinfo{volume}{123}},
  \bibinfo{pages}{184501} (\bibinfo{year}{2018}).

\bibitem[{\citenamefont{Khakifirooz et~al.}(2009)\citenamefont{Khakifirooz,
  Nayfeh, and Antoniadis}}]{khakifirooz2009simple}
\bibinfo{author}{\bibfnamefont{A.}~\bibnamefont{Khakifirooz}},
  \bibinfo{author}{\bibfnamefont{O.~M.} \bibnamefont{Nayfeh}},
  \bibnamefont{and}
  \bibinfo{author}{\bibfnamefont{D.}~\bibnamefont{Antoniadis}},
  \bibinfo{journal}{IEEE Transactions on Electron Devices}
  \textbf{\bibinfo{volume}{56}}, \bibinfo{pages}{1674} (\bibinfo{year}{2009}).

\bibitem[{\citenamefont{Ma and Jena}(2014)}]{ma2014charge}
\bibinfo{author}{\bibfnamefont{N.}~\bibnamefont{Ma}} \bibnamefont{and}
  \bibinfo{author}{\bibfnamefont{D.}~\bibnamefont{Jena}},
  \bibinfo{journal}{Physical Review X} \textbf{\bibinfo{volume}{4}},
  \bibinfo{pages}{011043} (\bibinfo{year}{2014}).

\bibitem[{\citenamefont{Radhakrishna}(2016)}]{radhakrishna2016modeling}
\bibinfo{author}{\bibfnamefont{U.}~\bibnamefont{Radhakrishna}}, Ph.D. thesis,
  \bibinfo{school}{Massachusetts Institute of Technology}
  (\bibinfo{year}{2016}).

\end{thebibliography}

\end{document}